\documentclass[preprintnumbers,amsmath,11pt,amssymb,floatfix,superscriptaddress,nofootinbib]{revtex4}
\usepackage{graphicx}
\usepackage{epsfig}
\usepackage{bm}
\usepackage{amsfonts}

\input epsf

\begin{document}

\title{Scalar Field Reconstructions of Standard, Power Law and Logarithmic Holographic Dark Energy with a Gauss-Bonnet IR cut-off}
\author{Antonio Pasqua}
\email{toto.pasqua@gmail.com}
\noaffiliation

\date{\today}
\newpage


\begin{abstract}
In this paper, we investigate the Holographic Dark Energy (HDE) model and its entropy-corrected versions, namely the Power Law and Logarithmic entropy corrected HDE models, by considering the infrared cut-off $L=\mathcal{G}^{-1/4}$, where $\mathcal{G}$ is the Gauss-Bonnet invariant. We derived the Equation of State parameter $\omega_D$, the deceleration parameter $q$ and the evolutionary form of the fractional energy density of DE $\Omega_D'$ for flat and non-flat universes, with and without interaction between DE and Dark Matter. We also analyzed the asymptotic behavior in the DE dominated epoch. Furthermore, correspondences between the considered HDE models and several scalar field models, including tachyon, k-essence, quintessence, Generalized Chaplygin Gas, Yang-Mills, and Nonlinear Electrodynamics models, were established.
\end{abstract}

\maketitle
\tableofcontents

\section{Introduction}
Observational cosmology strongly indicates that the present Universe is undergoing a phase of accelerated expansion. This picture is supported by a wide variety of independent astrophysical observations, including Type Ia Supernovae (SNeIa) measurements \cite{perlmutter,astier}, anisotropies in the Cosmic Microwave Background (CMB) radiation detected by the Wilkinson Microwave Anisotropy Probe (WMAP) satellite \cite{bennett-09-2003,spergel-09-2003}, studies of Large Scale Structure (LSS) formation \citep{tegmark,abz1,abz2}, as well as X-ray cluster observations \citep{allen}.

Within the standard Friedmann–Lemaitre-Robertson–Walker (FLRW) cosmological framework, this accelerated behavior is generally attributed to an exotic cosmic component characterized by negative pressure, commonly known as Dark Energy (DE). Current cosmological analyses, especially those based on WMAP observations \citep{spergel-09-2003,bennett-09-2003,peiris}, suggest that DE constitutes nearly two-thirds of the total energy content of the Universe, while Dark Matter (DM) accounts for most of the remaining fraction, with ordinary baryonic matter contributing only a small percentage.

Despite extensive observational support, the physical origin and nature of DE remain unresolved problems in modern cosmology. Numerous theoretical models have been proposed to explain this phenomenon (see \citep{Padmanabhan-07-2003,peebles,copeland-2006} for comprehensive reviews). The simplest and most widely studied candidate is the cosmological constant $\Lambda$, interpreted as vacuum energy with equation-of-state (EoS) parameter $\omega=-1$.

Although the $\Lambda$CDM scenario successfully explains many observational results, the cosmological constant is affected by two major theoretical difficulties: the fine-tuning problem and the cosmic coincidence problem \citep{copeland-2006}. The former originates from the enormous discrepancy between the observed vacuum energy density and the value predicted by quantum field theory, which differs by roughly $10^{-123}$ orders of magnitude \citep{weinberg}. The latter concerns the puzzling fact that the present energy densities of DE and DM are of the same order precisely in the current cosmological epoch.

These shortcomings have motivated the investigation of dynamical DE models in which the EoS parameter evolves with cosmic time. Observational analyses based on SNIa data indicate that time-dependent DE scenarios may provide a better description of the cosmic expansion history than a pure cosmological constant. In general, dynamical DE models can be classified into two broad categories. The first includes scalar-field constructions such as quintessence \citep{wetterich,ratra}, k-essence \citep{chiba-07-2000,armendariz-11-2000,armendariz-05-2001}, phantom models \citep{caldwell-10-2002,nojiri-06-2003,nojiri-07-2003}, tachyon fields \citep{sen-04-2002,Padmanabhan-06-2002,Padmanabhan-10-2002}, dilaton scenarios \citep{gasperini,piazza,arkani}, quintom cosmology \citep{elizalde-08-2004,nojiri-03-2005,anisimov}, and related approaches. The second category involves interacting or modified DE frameworks, including Chaplygin gas models \citep{Kamenshchik,setare-11-2007}, braneworld cosmologies \citep{deffayet,sahni}, Holographic Dark Energy (HDE) models \citep{cohen,horava-08-2000,setare-01-2007,setare-10-2007,setare-11-2006,setare-05-2007,setare-01-05-2007,setare-09-2007}, and Agegraphic Dark Energy (ADE) scenarios \citep{cai-12-2007,wei-02-2008}. A broad overview of DE models and related theoretical issues can be found in \citep{miao}.

An important conceptual development emerging from black hole physics and string theory is the holographic principle. This principle states that the number of physical degrees of freedom inside a given region should scale with the area of its boundary rather than with its volume \citep{thooft}. In addition, the total energy of a system must satisfy a constraint associated with an infrared (IR) cut-offf \citep{cohen}. Based on these ideas, Holographic Dark Energy was proposed as a viable DE candidate \citep{fischler}. Since then, HDE cosmology has attracted considerable attention and has been extensively explored in different contexts \citep{huang-08-2004,hsu,wang-09-2005,guberina-05-2005,guberina-05-2006,gong-09-2004,jamil-01-2010,J2011,
SJ2011,she2011,sheykhi-01-2011,sheykhi-03-01-2010,elizalde-05-2005,setare-02-2010,2011setare,2010setarejamil,2011karami,2010farooq,
SKJ2010,2010IJSF,2009JSS,karami-03-2010,sheykhi-11-2009,setare-09-2006,setare-08-2008,zhangX-08-2005,
zhangX-11-2006,zhangX-07-2007,enqvist-02-2005,shen,jamilfarooq2009}. Moreover, HDE models have been tested against several cosmological and astrophysical observations \citep{wangY,micheletti,zhangX-05-2009,feng-02-2005,kao}, and have also been discussed within the context of anthropic arguments \citep{huang-03-2005}.\\
The application of the holographic principle within a cosmological framework sets a maximum threshold for the cosmic entropy~\cite{fischler}. Adopting this approach, the energy density of the Holographic Dark Energy (HDE)  can be constrained as follows~\cite{li-12-2004}:
\begin{equation}
\rho_{D} \leq 3c^2 M_p^2 L^{-2}, \label{1}
\end{equation}
where $c^2$, $L$, and $M_p = (8\pi G)^{-1/2}$ represent a numerical constant, the IR cut-off and the reduced Planck mass, respectively. Here, the holographic bound is saturated when the equality is strictly met. In this setup, the definition of $\rho_{\Lambda}$ is intrinsically linked to the horizon entropy relation $S_{\mathrm{BH}} = A/(4G)$, where the area scales as $A \approx L^2$.\\
It is worth noting that a vast body of literature explores various modifications to this standard entropy expression, typically motivated by quantum-gravitational corrections or thermal fluctuations. In this work, we focus our attention on two widely studied extensions: the power law corrected entropy and the logarithmic corrected entropy frameworks.\\
The expression of the power law corrected entropy $S(A)_{\mathrm{pl}} $ is given by \citep{pl1,pl2}
\begin{eqnarray}
    S(A)_{\mathrm{pl}}  =   c_0 \left( \frac{A}{a_1^2} \right) \left[ 1 + c_1 f(A) \right],
    \label{powerlawentropyold}
\end{eqnarray}
where the functional form of $f(A)$ is specified as:
\begin{equation}
f(A) = \left( \frac{A}{a_1^2} \right)^{-\nu}, \label{powerlawentropyold2}
\end{equation}
thus introducing an explicit power law dependence on the horizon area $A$. Here, $c_0$ and $c_1$ represent dimensionless constants of order unity, $a_1$ denotes the UV cut-off scale at the horizon, and $\nu$ parameterizes the quantum mixing between ground and excited states. The horizon area is defined via the standard geometric relation $A=4\pi R_h^2$, where $R_h$ is the black hole event horizon radius. For macroscopic horizons (i.e. for $A \gg a_1^2$), the non-linear contribution from $f(A)$ becomes strictly negligible, allowing the entanglement entropy to asymptotically converge toward the classical Bekenstein-Hawking area law~\cite{Yassin:2020mjf,Tawfik:2015fda}. Alternatively, a highly operational formulation of this power law modified entropy can be written as:
\begin{equation}
S(A)_{\mathrm{pl}} = \frac{A}{4G} \left( 1 - K{\alpha} A^{1-\alpha/2} \right), \label{powerlawentropy}
\end{equation}
where $\alpha$ is a dimensionless parameter and the structural coefficient $K_{\alpha}$ is given by:
\begin{equation}K_{\alpha} = \frac{\alpha \left( 4\pi \right)^{\alpha/2-1}}{\left(4-\alpha \right)r^{2-\alpha}_c}, \label{murano1}
\end{equation}
with $r_c$ representing the infrared crossover scale. Assuming the underlying quantum wave function consists of a linear superposition of ground and excited states~\cite{pl3}, the sub-leading term in Eq.~\eqref{powerlawentropy} can be physically interpreted as an entanglement-induced power law correction to the area law. This corrective term scales prominently with higher excitations~\cite{pl3}. In the semi-classical macroscopic limit (large $A$), this correction decays rapidly, ensuring that the standard area law is identically recovered. Conversely, its impact becomes significant for microscopic black holes. Phenomenologically, this implies that for large horizon areas (low-energy regimes), the activation of high-energy modes is suppressed, leaving the ground-state configurations as the primary source of entanglement entropy. On the contrary, small horizons allow a substantial number of field modes to be excited, triggering substantial deviations from the linear area scaling. Motivated by the modified entropy relation of Eq.~\eqref{powerlawentropy}, the energy density of the Power Law Entropy Corrected HDE (PLECHDE) model can be readily derived as:
\begin{eqnarray}
\rho_{Dpl} &=& 3c^2 M_p^2 L^{-2} - \delta M_p^2 L^{-\gamma}\nonumber \\ 
&=& 3M_p^2 L^{-2} \left[ c^2 - \left(\frac{\delta}{3}\right) L^{-\gamma + 2} \right]. \label{newrhoold}
\end{eqnarray}
In the limiting case where $\delta = 0$, the PLECHDE density naturally reduces to the conventional HDE expression, $\rho_D = 3c^2 M_p^2 L^{-2}$. From an analytical perspective, the standard HDE framework can also be formally retrieved in the asymptotic limit $\gamma \rightarrow \infty$.\\
Beyond the power law framework, quantum-gravitational fluctuations and loop quantum gravity considerations suggest that the entropy-area relation can be alternatively modified as:
\citep{banerjee-04-2008,banerjee-06-2008,banerjee-05-2009,Majhi,wei-02-2010,easson,sad2010,jam2011}:
\begin{eqnarray}
    S_{BH} = \frac{A}{4G}+\tilde{\alpha} \log \left( \frac{A}{4G} \right) + \tilde{\beta},\label{2}
\end{eqnarray}
where the quantities  $\tilde{\alpha}$ and $\tilde{\beta}$ denote dimensionless coupling parameters. These specific logarithmic terms arise prominent in black hole thermodynamics when accounting for Loop Quantum Gravity (LQG) corrections, as well as from generic thermal fluctuations, microcanonical quantum corrections, or charge and mass variances. Notably, these quantum-gravitational modifications to the entropy-area scaling share a dual correspondence with curvature corrections integrated into the gravitational action~\cite{zhu,cai-08-2009,nojiri-2001}.\\
Employing this logarithmic-corrected entropy-area framework, the energy density characterizing the Logarithmic Entropy-Corrected HDE (LECHDE) model is given by~\cite{wei-10-2009}:
\begin{eqnarray}
    \rho_{D} =     3c^2 M_p^2 L^{-2} + \gamma_1 L^{-4} \log \left( M_p^2 L^2 \right)   + \gamma_2 L^{-4}, \label{3}
\end{eqnarray}
where the quantities $\gamma_1$ and $\gamma_2$ are dimensionless constants. In the limiting case where $\gamma_1 = \gamma_2 = 0$, Eq.~\eqref{3} naturally reduces to the standard HDE energy density. The leading term in Eq.~\eqref{3} represents the conventional holographic energy density component, whereas the second and third terms explicitly account for the logarithmic entropy corrections. Because these sub-leading corrections become comparable to the dominant term exclusively at small infrared cut-offf scales $L$, their evolutionary impact is predominantly confined to the early stages of the Universe. As the cosmic expansion proceeds and $L$ increases, these correction terms are progressively suppressed, and the total energy density smoothly tracks the standard HDE behavior.\\
In this work, we explore a HDE scenario where the IR cut-off is given by 
\begin{eqnarray}
    L=\mathcal{G}^{-1/4}\label{GBI},
\end{eqnarray}
where $\mathcal{G}$ represents the Gauss-Bonnet Invariant.\\
The Gauss-Bonnet invariant, usually denoted by $\mathcal{G}$, is defined in terms of the curvature tensors as
\begin{eqnarray}
\mathcal{G}
=
R^{2}
-4R_{\mu\nu}R^{\mu\nu}
+R_{\mu\nu\rho\sigma}R^{\mu\nu\rho\sigma},
\end{eqnarray}
where $R$ is the Ricci scalar, $R_{\mu\nu}$ is the Ricci tensor and $R_{\mu\nu\rho\sigma}$ is the Riemann curvature tensor. In a Friedmann–Lemaitre–Robertson–Walker (FLRW) spacetime, the Ricci scalar is given by
\begin{eqnarray}
    R=6\left( \dot H + 2H^2 + \frac{k}{a^2}  \right),
\end{eqnarray}
where $H=\frac{\dot a}{a}$ represents the Hubble parameter, $a(t)$ is the scale factor and $k$ is the curvature parameter which can assume the values $k = +1, 0, -1$ corresponding to a closed, a flat and an open Universe, respectively.\\ 
For a non-flat FLRW Universe, the Gauss-Bonnet invariant is given by
\begin{eqnarray}
\mathcal{G}=24\left(H^2 +\frac{k}{a^2}\right)\left(\dot{H} + H^2\right).
\label{17}
\end{eqnarray}
In the flat FLRW limit (i.e. for $k=0$), one obtains:
\begin{eqnarray}
\mathcal{G}=24H^2\left(\dot{H} + H^2\right).
\label{17flat}
\end{eqnarray}
By employing the relation
\begin{eqnarray}
\frac{\ddot{a}}{a}
= \dot{H}+H^{2},
\end{eqnarray}
the Gauss-Bonnet invariant $\mathcal{G}$ can alternatively be written as:
\begin{eqnarray}
\mathcal{G}
&=&
24
\left(\frac{\ddot{a}}{a}\right)
\left[
\left(\frac{\dot{a}}{a}\right)^2
+
\frac{k}{a^{2}}
\right],
\end{eqnarray}
for the non flat case, while it can be written as:
\begin{eqnarray}
\mathcal{G}
&=&
24\left(\frac{\dot{a}}{a}\right)^2\left(
\frac{\ddot{a}}{a}\right)
\end{eqnarray}
for the flat case.\\
One of the main motivations for considering the Gauss-Bonnet Invariant as IR cut-off is that the Gauss-Bonnet term frequently emerges in several approaches to quantum gravity and string-inspired cosmology. In particular, it appears as a higher-curvature correction in the low-energy effective action of string theory and in modified gravity scenarios. Therefore, using $\mathcal{G}$ as IR cut-off may offer a deeper geometrical interpretation of the HDE models and may encode quantum gravitational effects more naturally than phenomenological cut-offs based exclusively on horizon distances.

Another important advantage of the Gauss--Bonnet cut-off is that it depends only on local geometric quantities. In contrast to the future event horizon cut-off, which suffers from causality issues because it depends on the future evolution of the Universe, the Gauss-Bonnet invariant is completely determined by local cosmological dynamics. This makes the model theoretically more appealing and avoids the non-local character of event horizon based HDE models.

From a cosmological perspective, the Gauss-Bonnet HDE model may exhibit significant differences compared to Ricci HDE model: for this mode, the IR cut-off is associated with the Ricci scalar $R$, which depends linearly on $H$ and $\dot H$. By contrast, the Gauss-Bonnet Invariant contains quadratic contributions in the Hubble parameter, leading to a stronger nonlinear coupling between the DE density and the cosmic expansion dynamics. As a consequence, the cosmological evolution of the EoS parameter, the deceleration parameter $q$ and the DE fractional energy density can differ substantially from the Ricci HDE scenario, particularly during the transition from matter domination to accelerated expansion.

Moreover, the Gauss-Bonnet cut-off can generate richer dynamical behaviors, including modified phantom-like evolution, different stability properties, and distinct asymptotic DE Dominated regimes. Since the invariant vanishes in certain cosmological phases and becomes dominant at high curvature scales, it may also provide a natural connection between early-Universe dynamics and late-time cosmic acceleration.
Using Eq. (\ref{GBI}) in the expressions of the energy densities defined in Eqs. (\ref{1}), (\ref{newrhoold}) and (\ref{3}), we obtain the following results:
\begin{eqnarray}
    \rho_{D}&=&3c^2M_p^2\,\sqrt{\mathcal{G}},
    \label{vialli1}\\
    \rho_{Dpl}&=&  3M_{p}^2 \sqrt{\mathcal{G}} \left[ c^2 - \left(\frac{\delta}{3}\right) \mathcal{G}^{\frac{\gamma-2}{4}} \right], \label{newrhooldG}\\\
\rho_{Dlog} &=& 3c^2 M_p^2 \sqrt{\mathcal{G}} + \gamma_1 \mathcal{G}\log \left( \frac{M_p^2}{\sqrt{\mathcal{G}}} \right) + \gamma_2 \mathcal{G}. \label{vialli2}
\end{eqnarray}
This paper is organized in the following way In Section 2,we outline the physical context of our framework and systematically derive the explicit expressions for the Equation of State (EoS) parameter of DE $\omega_{D}$, the deceleration parameter $q$ and the evolutionary form of the DE density parameter, denoted by $\Omega_D'$     by considering the following four scenarios:
\begin{enumerate}
    \item flat non-interacting Universe,
    \item non-flat non-interacting Universe,
    \item flat interacting Universe,
    \item non-flat interacting Universe.
\end{enumerate} 
In Section 3, we establish a correspondence between our
models and some scalar fields model, in particular the Tachyon, the k-essence, the Quintessence, the Generalized Chaplygin Gas (GCG), the Yang-Mills (YM) and the Nonlinear Electrodynamics (NLED) fields. Finally, in Section 4 we write
the Conclusions. 

\section{Cosmological Framework}
We construct our cosmological model on the foundation of a non-flat Friedmann-Lemaître-Robertson-Walker (FLRW) geometry. The background spacetime is characterized by the following invariant line element:
\begin{equation}
ds^2 = -dt^2 + a^2(t) \left[ \frac{dr^2}{1-kr^2} + r^2 \left(d\theta^2 + \sin^2\theta d\varphi^2 \right) \right], \label{6}
\end{equation}
Probing the gravitational sector via the Einstein field equations yields the first Friedmann equation:
\begin{equation}
H^2 + \frac{k}{a^2} = \frac{1}{3M^2_p} \left( \rho_{D} + \rho_{m} \right), \label{7}
\end{equation}
where  $\rho_{D}$ and $\rho_{m}$ account for the energy densities of DE and Dark Matter (DM) 
components, respectively. By normalizing Eq.~\eqref{7} against the critical density threshold, defined as $\rho_{cr} \equiv 3M^2_pH^2$, we can introduce the set of dimensionless density parameters for DM, curvature and DE:
\begin{eqnarray}
\Omega_m &=& \frac{\rho_m}{3M_p^2H^2},\label{8-10-1}\\
\Omega_k &=& -\frac{k}{H^2a^2},\label{8-10-2}\\
\Omega_{D} &=& \frac{\rho_{D}}{\rho_{3M_p^2H^2}}. \label{8-10-3}
\end{eqnarray}
While global spatial flatness remains a core assumption in baseline models, contemporary CMB measurements do not rule out a closed topology, indicating a marginal positive curvature contribution near $\Omega_k \approx 0.02$~\cite{spergel-06-2007}. Under these definitions, the geometric constraint of Eq.~\eqref{7} simplifies to the conservation law:
\begin{equation}
\Omega_{m} + \Omega_{D} = 1 + \Omega_k. \label{11}
\end{equation}
Covariant conservation of the stress-energy tensor ($\nabla_{\mu}T^{\mu\nu}=0$) requires the combined dark sector fluid, $\rho_{\mathrm{tot}} = \rho_{D} + \rho_m$, to satisfy the fundamental continuity equation:
\begin{equation}
\dot{\rho}_{tot} + 3H(1 + \omega)\rho_{tot} = 0, \label{12}
\end{equation}
where the parameter $\omega = p_{\mathrm{tot}}/\rho_{\mathrm{tot}}$ establishes the effective cosmic Equation of State (EoS) parameter.
If there is not interaction between DE and DM, Eq. (\ref{12}) leads to the following continuity equations for DE and DM:
\begin{eqnarray}
\dot{\rho}_{D} + 3H\rho_{D}(1 + \omega_{D}) &=& 0, \label{13Non} \\
\dot{\rho}_m + 3H\rho_m &=& 0. \label{14Non}
\end{eqnarray}
To account for a possible coupled dynamics between the dark components, we introduce a phenomenological energy exchange channel, $Q$. This term explicitly breaks the independent conservation of the two fluids, yielding the coupled differential system:
\begin{eqnarray}
\dot{\rho}_{D} + 3H\rho_{D}(1 + \omega_{D}) &=& -Q, \label{13} \\
\dot{\rho}_m + 3H\rho_m &=& Q. \label{14}
\end{eqnarray}
The source term $Q$ is in general a function of the energy densities of DE and DM: 
\begin{equation}
Q = 3b^2Hf(\rho_m , \rho_{D}), \label{15}
\end{equation}
where $b^2$ represents a dimensionless cross-coupling parameter and $f\left(\rho_m,\rho_D \right)$ is a generic function of $\rho_m$ and $\rho_D$. Within this framework, a positive coupling ($b^2 > 0$) governs a continuous energy flow from DE into the DM sector. This specific dynamic aligns with observational signatures identified in macroscopic systems like the Abell A586 cluster~\cite{bertolami-10-2007,jamil-11-2008}. Conversely, negative values of $b^2$ are avoided to prevent unphysical back-reaction processes that violate the second law of thermodynamics. Joint cosmological constraints from CMB data and cluster abundances place a strict upper limit on this parameter, restricting it to the weak-coupling domain ($b^2 \lesssim 0.025$)~\cite{ichiki,amendola-04-2007}. This ensures that the interaction induces only a subtle modification to the standard background evolution without disrupting the well-tested matter-dominated era.

We now derive the main cosmological quantities we want to derive, namely the DE Equation of State (EoS) parameter $\omega_D$, the evolution equation of the DE fractional density $\Omega_D'$, and the deceleration parameter $q$.\\
We start considering the flat non-interacting Universe.\\
From the Friedmann equation (\ref{7}), one obtains
\begin{eqnarray}
    \dot{H}=-\frac{1}{2M_p^2}\left[\rho_m+\rho_{D}\left(1+\omega_{D}\right)\right].
    \label{18}
\end{eqnarray}
By adding Eqs. (\ref{7}) and (\ref{18}), we find
\begin{eqnarray}
    \dot{H}+H^2
    &=&
    \frac{\rho_m+\rho_{D}}{3M_p^2}
    -\frac{1}{2M_p^2}
    \left[
    \rho_m+\rho_{D}\left(1+\omega_{D}\right)
    \right]
    \nonumber\\
    &=&
    -\frac{1}{6M_p^2}
    \left[
    \rho_m+\rho_{D}\left(1+3\omega_{D}\right)
    \right].
    \label{19}
\end{eqnarray}
Therefore, the Gauss-Bonnet invariant $\mathcal{G}$ can be expressed as
\begin{eqnarray}
\mathcal{G}
=
-\frac{4}{3M_p^4}
\left(\rho_D+\rho_m\right)
\left[
\rho_m+\rho_D\left(1+3\omega_D\right)
\right].
\label{20}
\end{eqnarray}
From Eq. (\ref{20}), we can straightforwardly derive the expression for the DE EoS parameter $\omega_D$:
\begin{equation}
    \omega_{D}
    =
    -\frac{\rho_{D}}{36c^4(\rho_D + \rho_m)}
    - \frac{\rho_m}{3\rho_D}
    - \frac{1}{3}.
    \label{sinner2}
\end{equation}

Using the fractional energy densities, Eq. (\ref{sinner2}) can be rewritten as
\begin{eqnarray}
    \omega_{D}
    &=&
    -\frac{\Omega_{D}}{36c^4(\Omega_D + \Omega_m)}
    - \frac{\Omega_m}{3\Omega_D}
    - \frac{1}{3}
    \nonumber \\
    &=&
    -\frac{\Omega_{D}}{36c^4}
    - \frac{1-\Omega_D}{3\Omega_D}
    - \frac{1}{3}
    \nonumber \\
    &=&
    -\frac{\Omega_{D}}{36c^4}
    - \frac{1}{3\Omega_D}.\label{alcatraz}
\end{eqnarray}

From the continuity equation for DE given in Eq. (\ref{13Non}), we obtain
\begin{eqnarray}
    \dot{\rho}_{D} =
    3H    \left( -\rho_{D} -\rho_{D}\omega_{D} \right).
    \label{23nonI}
\end{eqnarray}

Substituting the expression of $\omega_D$ from Eq. (\ref{sinner2}) into Eq. (\ref{23nonI}), we obtain
\begin{eqnarray}
\dot{\rho}_{D}
= 3H
\left[ \frac{\rho_m}{3}
+ \frac{\rho_{D}^2}{36c^4(\rho_D + \rho_m)}
- \frac{2\rho_D}{3}
\right].
\label{23nonI-2}
\end{eqnarray}

Dividing by the critical density $\rho_{cr}=3H^2M_p^2$, Eq. (\ref{23nonI-2}) becomes
\begin{eqnarray}
    \frac{\dot{\rho}_{D}}{\rho_{cr}}
    &=&  \dot{\Omega}_{D}    +    2\Omega_{D}\left(\frac{\dot{H}}{H}\right)
    \nonumber\\
    &=&    3H\left[    \frac{\Omega_m}{3}
    +    \left(\frac{\rho_D}{\rho_{cr}}
    \right)     \frac{\rho_{D}}{36c^4(\rho_D + \rho_m)}
    -    \frac{2\Omega_D}{3} \right]
    \nonumber \\
    &=&
    3H    \left\{ \frac{1-\Omega_D}{3}
    +    \Omega_D\left[    \frac{\Omega_D}{36c^4(\Omega_D + \Omega_m)}
    \right]    -    \frac{2\Omega_D}{3}
    \right\}    \nonumber \\
    &=&
    3H    \left[\frac{1-\Omega_D}{3}+    \frac{\Omega_D^2}{36c^4}
    -     \frac{2\Omega_D}{3} \right]
    \nonumber \\
    &=&
    3H  \left[\frac{1}{3}-\Omega_D    + \frac{\Omega_D^2}{36c^4}\right].
    \label{25}
\end{eqnarray}

Therefore, we obtain that:
\begin{eqnarray}
  \dot{\Omega}_{D} &=& 2\Omega_{D}
    \left(\frac{\dot{H}}{H} \right)
    +    3H\left[\frac{1}{3}-\Omega_D
    +    \frac{\Omega_D^2}{36c^4}\right].
    \label{26}
\end{eqnarray}
Since   
\begin{eqnarray}
    \Omega_{D}'=\frac{d\Omega_{D}}{dx}= \left(\frac{1}{H}\right)\dot{\Omega}_{D},
\end{eqnarray}
where $x=\ln a$, we derive:
\begin{eqnarray}
\Omega_{D}'
    &=& -2\Omega_{D}\left(\frac{\dot{H}}{H^2}\right)+\left[1 -3\Omega_D   +  \frac{\Omega_D^2}{12c^4}     \right].  \label{25}
\end{eqnarray}
Using the relation:
\begin{equation}
\dot{H} = \frac{dH}{dt} = \frac{dx}{dt} \frac{dH}{dx} = H H',
\end{equation}
Eq. (\ref{25}) can be rewritten as
\begin{eqnarray}
\Omega_{D}'
    &=& -2\Omega_{D}\left(\frac{H' }{H}\right)+\left[1 -3\Omega_D   +  \frac{\Omega_D^2}{12c^4}     \right]\nonumber \\
    &=& -\Omega_{D}\left[ 3+  2 \left(\frac{H' }{H}\right)\right]+1    +  \frac{\Omega_D^2}{12c^4}    .\label{25-}
\end{eqnarray}
Next, we derive an explicit expression for $\frac{H'}{H}$. The deceleration parameter $q$ is defined as
\begin{eqnarray}
    q= -1 - \frac{\dot H}{H^2} = -1-\frac{H'}{H}.
\end{eqnarray}
Moreover, for a spatially flat Universe, one has
\begin{eqnarray}
    q=\frac{1}{2}\left(1  + 3\Omega_{D} \omega_{D}  \right).\label{32flat}
\end{eqnarray}
By equating the above expressions for $q$, we obtain
\begin{eqnarray}
    -1-\frac{H'}{H}=\frac{1}{2}\left(1  + 3\Omega_{D} \omega_{D}  \right)\label{},
\end{eqnarray}
which leads to
\begin{eqnarray}
   \frac{H'}{H}=-\frac{3}{2}\left(1  + \Omega_{D} \omega_{D}  \right).\label{}
\end{eqnarray}
Using the expression for $\omega_D$ derived in Eq. (\ref{alcatraz}), we finally obtain $\frac{H'}{H}$ as:
\begin{eqnarray}
\frac{H'}{H}=-1  + \frac{\Omega_{D}^2}{24c^4}.\label{sinner3}
\end{eqnarray}
Inserting Eq. (\ref{sinner3}) into Eq. (\ref{25-}), we obtain that $\Omega_D'$ can be written as
\begin{equation}
    \Omega_{D}' = (1 - \Omega_D) \left( 1 + \frac{\Omega_D^2}{12c^4} \right) .\label{sinner4}
\end{equation}
Eq. (\ref{sinner4}) does not admit a general analytical solution. However, in the limiting case $12c^2 = \Omega_D$, it reduces to
\begin{equation}
    \Omega_{D}' = (1 - \Omega_D) \left( 1 + \Omega_D\right) = 1-\Omega_D^2.
\end{equation}
In this case, the differential equation admits the exact solution
\begin{eqnarray}
    \Omega_D(x) = \tanh (x+C),
\end{eqnarray}
where $C$ is an integration constant.\\
We now compute the cosmological quantities in the limiting case of a spatially flat, Dark Energy dominated Universe, i.e. $\Omega_k=\Omega_m=0$ and $\Omega_D=1$.\\
In this case, the Friedmann equation reduces to
\begin{eqnarray}
    H^2 = \rho_D= 3c^2M_p^2\sqrt{\mathcal{G}} =  3c^2M_p^2\sqrt{24}H\left( H^2 + \dot H\right)^{1/2}.
\end{eqnarray}
This expression is equivalent to:
\begin{eqnarray}
   \dot H = -H^2 \left( 1 - \frac{1}{216c^4M_p^4} \right) \label{sinner5},
\end{eqnarray}
whose solution is given by:
\begin{equation}
    H(t) = \frac{1}{\left( 1 - \frac{1}{216 c^4 M_p^4} \right) t + \frac{1}{H_0}},
\end{equation}
where $H_0$ denotes the present day value of the Hubble parameter $H$.\\
The scale factor $a(t)$ is then obtained as
\begin{equation}
    a(t) = a_0 \left[ 1 + \left( 1 - \frac{1}{216 c^4 M_p^4} \right) H_0 t \right]^{\frac{1}{1 - \frac{1}{216 c^4 M_p^4}}}.
\end{equation}
The DE density $\rho_D$ is given by
\begin{eqnarray}
    \rho_D = H^2 = \left[ \frac{1}{\left( 1 - \frac{1}{216 c^4 M_p^4} \right) t + \frac{1}{H_0}}\right]^2.
\end{eqnarray}
Finally, the EoS and deceleration parameters read
\begin{eqnarray}
w_D &=& -\frac{1}{3} - \frac{1}{324c^4 M_p^4},\\
q &=& -\frac{1}{216c^4 M_p^4}.
\end{eqnarray}
Thus, both $w_D$ and $q$ are constant negative parameters in this regime.

Following the same procedure, we now extend the analysis to the case of a non-flat Universe.\\
Starting from the expression
\begin{equation}
    \omega_{D} = -\frac{\rho_{D}}{36c^4(\rho_D + \rho_m)} - \frac{\rho_m}{3\rho_D} - \frac{1}{3},
\end{equation}
we obtain the general expression for the DE EoS parameter in the presence of spatial curvature:
\begin{eqnarray}
    \omega_{D,k} &=& -\frac{\Omega_{D}}{36c^4(\Omega_D + \Omega_m)} - \frac{\Omega_m}{3\Omega_D} - \frac{1}{3}\nonumber \\
    &=&-\frac{\Omega_{D}}{36c^4(1 + \Omega_k)} - \frac{1-\Omega_D+\Omega_k }{3\Omega_D} - \frac{1}{3}\nonumber \\
    &=&-\frac{\Omega_{D}}{36c^4(1 + \Omega_k)} -\frac{1+\Omega_k }{3\Omega_D}.
\end{eqnarray}
The evolution equation of the DE fractional density becomes
\begin{eqnarray}
\Omega_{D,k}' &=& (1 - \Omega_D) \left[ 1 + \Omega_k + \frac{\Omega_D^2}{12c^4(1+\Omega_k)} \right]. \label{eq:Omega_D_final}
\end{eqnarray}
In the limiting case of $\Omega_k = 0$, the previous result is recovered.\\
The general expression of the deceleration parameter $ q$ reads
\begin{eqnarray}
    q=\frac{1}{2}\left(1  +\Omega_k+ 3\Omega_{D} \omega_{D}  \right).\label{32flat}
\end{eqnarray}

Finally, the evolution equation for $\Omega_D$ in the presence of interaction  takes the form
\begin{eqnarray}
\Omega_{D,int}' &=& (1 - \Omega_D) \left( 1 + \frac{\Omega_D^2}{12c^4} \right) -\frac{Q}{H\rho_{cr}},
\end{eqnarray}
while in presence of interaction and curvature takes the form:
\begin{eqnarray}
\Omega_{D,int,k}' &=& (1 - \Omega_D) \left[ 1 + \Omega_k + \frac{\Omega_D^2}{12c^4(1+\Omega_k)} \right] -\frac{Q}{H\rho_{cr}}.
\end{eqnarray}
We now derive the general expression for $\omega_D$ as function of $\rho_{Dpl}$ and $\rho_{Dlog}$.

From Eq. (\ref{20}), we obtain that the EoS parameter $\omega_D$ can be written as
\begin{eqnarray}    
\omega_{D} = - \frac{M_p^4 \mathcal{G}}{4\rho_{D}(\rho_m + \rho_{D})} - \frac{\rho_m + \rho_{D}}{3\rho_{D}}.\label{21}
\end{eqnarray}
Rewriting the above expression of $\omega_D$ in terms of fractional energy densities, we obtain
\begin{eqnarray}    
\omega_{D} &=& - \frac{M_p^4 \mathcal{G}}{4\rho_{D}(\rho_m + \rho_{D})} - \frac{\rho_m + \rho_{D}}{3\rho_{D}}\nonumber \\
&=&-\frac{M_p^4 \mathcal{G}}{4\rho_{D}^2}\frac{\rho_{D}}{\rho_m + \rho_{D}} - \frac{\rho_m + \rho_{D}}{3\rho_{D}}\nonumber \\
&=&-\frac{M_p^4 \mathcal{G}}{4\rho_{D}^2}\frac{\Omega_{D}}{\Omega_m + \Omega_{D}} - \frac{\Omega_m + \Omega_{D}}{3\Omega_{D}}\nonumber \\
&=&-\Omega_{D}\left(\frac{M_p^4 \mathcal{G}}{4\rho_{D}^2}\right)- \frac{1 }{3\Omega_{D}}
.\label{21.new}
\end{eqnarray}
Therefore, we obtain the following expressions for the EoS parameters for the PLECHDE and LECHDE models:
\begin{eqnarray}
    \omega_{Dpl} &=&-\Omega_{D}\left(\frac{M_p^4 \mathcal{G}}{4\rho_{Dpl}^2}\right)- \frac{1 }{3\Omega_{D}},\\
    \omega_{Dlog} &=&-\Omega_{D}\left(\frac{M_p^4 \mathcal{G}}{4\rho_{Dlog}^2}\right)- \frac{1 }{3\Omega_{D}}.
\end{eqnarray}
In the presence of curvature, the EoS parameter becomes
\begin{eqnarray}    
\omega_{D} &=& - \frac{M_p^4 \mathcal{G}}{4\rho_{D}(\rho_m + \rho_{D})} - \frac{\rho_m + \rho_{D}}{3\rho_{D}}\nonumber \\
&=&-\frac{M_p^4 \mathcal{G}}{4\rho_{D}^2}\frac{\rho_{D}}{\rho_m + \rho_{D}} - \frac{\rho_m + \rho_{D}}{3\rho_{D}}\nonumber \\
&=&-\frac{M_p^4 \mathcal{G}}{4\rho_{D}^2}\frac{\Omega_{D}}{\Omega_m + \Omega_{D}} - \frac{\Omega_m + \Omega_{D}}{3\Omega_{D}}\nonumber \\
&=&-\frac{\Omega_{D}}{1+\Omega_k}\left(\frac{M_p^4 \mathcal{G}}{4\rho_{D}^2}\right)- \frac{1+\Omega_k }{3\Omega_{D}}
.\label{21.newK}
\end{eqnarray}
Thus, we obtain the following expressions for $\omega_{Dpl}$ and $\omega_{Dlog}$:
\begin{eqnarray}    
\omega_{Dpl,k}
&=&-\frac{\Omega_{D}}{1+\Omega_k}\left(\frac{M_p^4 \mathcal{G}}{4\rho_{Dpl}^2}\right)- \frac{1+\Omega_k }{3\Omega_{D}},
\label{}\\
\omega_{Dlog,k}
&=&-\frac{\Omega_{D}}{1+\Omega_k}\left(\frac{M_p^4 \mathcal{G}}{4\rho_{Dlog}^2}\right)- \frac{1+\Omega_k }{3\Omega_{D}}.
\end{eqnarray}
In the limiting case corresponding to $\Omega_k = 0$, the previous results are recovered.

Using the expressions of $\rho_{Dpl}$ and $\rho_{Dlog}$, we obtain the following expression for $\frac{\mathcal{G}}{\rho_{Dpl}^2}$ and $\frac{\mathcal{G}}{\rho_{Dlog}^2}$:
\begin{eqnarray}
\frac{\mathcal{G}}{\rho_{Dpl}^2} &=& \frac{1}{M_p^4 \left( 3c^2 - \delta \mathcal{G}^{\frac{\gamma - 2}{4}} \right)^2},\\
\frac{\mathcal{G}}{\rho_{Dlog}^2} &=& \frac{1}{\mathcal{G} \left[ \frac{3c^2 M_p^2}{\sqrt{\mathcal{G}}} + \gamma_1 \log \left( \frac{M_p^2}{\sqrt{\mathcal{G}}} \right) + \gamma_2 \right]^2}.
\end{eqnarray}

\section{Correspondence with Scalar Field Models}
We now establish an explicit correspondence between the DE models we studied and several well-known scalar fields models, in particular the Tachyon, the k-essence, the Quintessence, the Generalized Chaplygin Gas (GCG), the Yang-Mills (YM) and the Nonlinear Electrodynamics (NLED) fields. Such correspondences are particularly valuable because these specific formulations act as effective field representations of a more fundamental, microscopic Dark Energy theory.\\
For this reason, we first compare the energy density of DE models we consider in this paper with the energy density of the corresponding scalar field model under investigation. 
Subsequently, we equate the EoS parameters of the scalar field models with the EoS parameters of our DE models.
For other works of reconstruction of scalar field models, see \cite{mioS1,mioS2,mioS3,mioS4,mioS5,mioS6,mioS7}

\subsection{Tachyon Scalar Field Model}
We start with the first scalar field model considered in this paper, the tachyon.\\
In recent years, tachyon fields have attracted considerable attention in the context of early-Universe inflation and late-time cosmic acceleration, as they can also provide a viable candidate for DE \citep{bagla, shao, calcagni, copeland-02-2005}.

In string theory, the tachyon is interpreted as an unstable scalar degree of freedom that emerges naturally in the Dirac-Born-Infeld (DBI) formulation describing the dynamics of D-branes
\citep{sen-07-2002,sen-10-1999,bergshoeff,klusovn,kutasov}. Its condensation mechanism near the maximum of the effective potential makes it a natural candidate for driving an inflationary phase in the early Universe.

One of the most interesting features of tachyon cosmology is that the corresponding rolling field exhibits an EoS parameter that evolves continuously from $\omega=-1$ to $\omega=0$. This behaviour has motivated the interpretation of dark energy as a dynamical component rather than a strict cosmological constant, and has led to several inflationary scenarios based on tachyon dynamics.

The effective Lagrangian of the tachyon field is given by:
\begin{eqnarray}
L=-V(\phi)\sqrt{1-g^{\mu \nu}\partial _{\mu}\phi \partial_{\nu}\phi},\label{39}
\end{eqnarray}
where $V(\phi)$ denotes the tachyon potential and $g^{\mu\nu}$ is the spacetime metric tensor. The energy density $\rho_{\phi}$ and pressure $p_{\phi}$ of the tachyon field are given, respectively, by
\begin{eqnarray}
    \rho_{\phi}&=&\frac{V(\phi)}{\sqrt{1-\dot{\phi}^2}},\label{40}\\
p_{\phi}&=& -V(\phi)\sqrt{1-\dot{\phi}^2}.\label{41}
\end{eqnarray}
The corresponding EoS parameter of tachyon scalar field model $\omega_{\phi}$ is therefore given by:
\begin{eqnarray}
\omega_{\phi}=\frac{p_{\phi}}{\rho_{\phi}}=\dot{\phi}^2-1.\label{42}
\end{eqnarray}
To ensure a real and well-defined energy density, the condition $0 \leq \dot{\phi}^2 < 1$ must hold. Consequently, the EoS parameter is constrained to lie in the range $-1 < \omega_{\phi} < 0$. This implies that the tachyon field can drive accelerated expansion, but it cannot cross into the phantom regime ($\omega < -1$).

By comparing the energy density of our models $\omega_D$ and Eq. (\ref{40}), the tachyon potential $V(\phi)$ can be expressed as:
\begin{eqnarray}
    V(\phi)=\rho_{D} \sqrt{1-\dot{\phi}^2}.\label{43}
\end{eqnarray}
Instead, equating the energy density of our models $\omega_D$ and Eq.  (\ref{42}), one obtains an expression for the kinetic term $\dot{\phi}^2$ given by:
\begin{eqnarray}
\dot{\phi}^2_{D}&=& 1 +\omega_{D} \nonumber \\
&=&1-\frac{\Omega_{D}}{36c^4} - \frac{1}{3\Omega_D},\label{44}\\
\dot{\phi}^2_{D,k}&=& 1 +\omega_{D,k} \nonumber \\
&=&1-\frac{\Omega_{D}}{36c^4(1 + \Omega_k)} -\frac{1+\Omega_k }{3\Omega_D},\label{44-1}\\
\dot{\phi}^2_{Dpl}&=& 1 +\omega_{Dpl} \nonumber \\
&=&1-\Omega_{D}\left(\frac{M_p^4 \mathcal{G}}{4\rho_{Dpl}^2}\right)- \frac{1 }{3\Omega_{D}},\label{44-2}\\
\dot{\phi}^2_{Dlog}&=& 1 +\omega_{Dlog} \nonumber \\
&=&1-\Omega_{D}\left(\frac{M_p^4 \mathcal{G}}{4\rho_{Dlog}^2}\right)- \frac{1 }{3\Omega_{D}},\label{44-3}\\
\dot{\phi}^2_{Dpl,k}&=& 1 +\omega_{Dpl,k} \nonumber \\
&=&1-\frac{\Omega_{D}}{1+\Omega_k}\left(\frac{M_p^4 \mathcal{G}}{4\rho_{Dpl}^2}\right)- \frac{1+\Omega_k }{3\Omega_{D}},\label{44-4}\\
\dot{\phi}^2_{Dlog,k}&=& 1 +\omega_{Dlog,k} \nonumber \\
&=&1-\frac{\Omega_{D}}{1+\Omega_k}\left(\frac{M_p^4 \mathcal{G}}{4\rho_{Dlog}^2}\right)- \frac{1+\Omega_k }{3\Omega_{D}}\label{44-5}.
\end{eqnarray}
Using the results we obtained for the DE models we study, the tachyon potential $V(\phi)$ can be written as:
\begin{eqnarray}
    V\left( \phi  \right)_{D} &=& \rho _{D} \sqrt{-\omega_{D}}\nonumber    \\
    &=& \rho _{D} \sqrt{  \frac{\Omega_{D}}{36c^4} + \frac{1}{3\Omega_D}  },\label{} \\
    V\left( \phi  \right)_{D,k} &=& \rho _{D,k} \sqrt{-\omega_{D,k} }\nonumber    \\
    &=& \rho _{D,k} \sqrt{ \frac{\Omega_{D}}{36c^4(1 + \Omega_k)} +\frac{1+\Omega_k }{3\Omega_D}   },\label{} \\
    V\left( \phi  \right)_{Dpl} &=& \rho _{Dpl} \sqrt{-\omega_{Dpl}}\nonumber    \\
    &=& \rho _{Dpl} \sqrt{ \Omega_{D}\left(\frac{M_p^4 \mathcal{G}}{4\rho_{Dpl}^2}\right)+ \frac{1 }{3\Omega_{D}}   },\label{} \\
    V\left( \phi  \right)_{Dlog} &=& \rho _{Dlog} \sqrt{-\omega_{Dlog}}\nonumber    \\
    &=& \rho _{Dlog} \sqrt{  \Omega_{D}\left(\frac{M_p^4 \mathcal{G}}{4\rho_{Dlog}^2}\right)+\frac{1 }{3\Omega_{D}}  },\label{} \\
    V\left( \phi  \right)_{Dpl,k} &=& \rho _{Dpl,k} \sqrt{-\omega_{Dpl,k}}\nonumber    \\
    &=& \rho _{Dpl,k} \sqrt{  \frac{\Omega_{D}}{1+\Omega_k}\left(\frac{M_p^4 \mathcal{G}}{4\rho_{Dpl}^2}\right)+ \frac{1+\Omega_k }{3\Omega_{D}}  },\label{} \\
    V\left( \phi  \right)_{Dlog,k} &=& \rho _{Dlog,k} \sqrt{-\omega_{Dlog,k}}\nonumber    \\
    &=& \rho _{Dlog,k} \sqrt{  \frac{\Omega_{D}}{1+\Omega_k}\left(\frac{M_p^4 \mathcal{G}}{4\rho_{Dlog}^2}\right)+ \frac{1+\Omega_k }{3\Omega_{D}}  }\label{} .
\end{eqnarray}
We obtain from above equations that the kinetic energy term $\dot{\phi}^2$ and the potential $V(\phi)$ are well-defined only if the following condition is satisfied:
\begin{eqnarray}
    -1\leq \omega_{D} \leq 0,\label{46}
\end{eqnarray}
which means that the phantom divide cannot be crossed in a universe undergoing accelerated expansion.\\
If we consider the relation $\dot{\phi}=\phi'H$, we derive:
\begin{eqnarray}
 \phi'_{D}&=& \frac{1}{H}\cdot  \sqrt{   1-\frac{\Omega_{D}}{36c^4(1 + \Omega_k)} -\frac{1+\Omega_k }{3\Omega_D}  },\\ 
  \phi'_{D,k}&=& \frac{1}{H}\cdot  \sqrt{  1-\frac{\Omega_{D}}{36c^4(1 + \Omega_k)} -\frac{1+\Omega_k }{3\Omega_D}   },\\ 
  \phi'_{Dpl}&=& \frac{1}{H}\cdot  \sqrt{   1-\Omega_{D}\left(\frac{M_p^4 \mathcal{G}}{4\rho_{Dpl}^2}\right)- \frac{1 }{3\Omega_{D}}  },\\ 
  \phi'_{Dlog}&=& \frac{1}{H}\cdot \sqrt{ 1-\Omega_{D}\left(\frac{M_p^4 \mathcal{G}}{4\rho_{Dlog}^2}\right)- \frac{1 }{3\Omega_{D}}    },\\ 
  \phi'_{Dpl,k}&=& \frac{1}{H}\cdot    \sqrt{  1-\frac{\Omega_{D}}{1+\Omega_k}\left(\frac{M_p^4 \mathcal{G}}{4\rho_{Dpl}^2}\right)- \frac{1+\Omega_k }{3\Omega_{D}}   },\\ 
  \phi'_{Dlog,k}&=& \frac{1}{H}\cdot  \sqrt{  1-\frac{\Omega_{D}}{1+\Omega_k}\left(\frac{M_p^4 \mathcal{G}}{4\rho_{Dlog}^2}\right)- \frac{1+\Omega_k }{3\Omega_{D}}   }.
\end{eqnarray}
Accordingly, one can deduce the evolutionary behavior of the tachyon scalar field in the following form:
\begin{eqnarray}
    \phi\left(a\right)_{D} - \phi\left(a_0\right)_{D}&=&\int_{a_0}^a \frac{da}{aH}\times\nonumber \label{}\\
    &&\sqrt{1-\frac{\Omega_{D}}{36c^4(1 + \Omega_k)} -\frac{1+\Omega_k }{3\Omega_D} },\\
    \phi\left(a\right)_{D,k} - \phi\left(a_0\right)_{D,k}&=&\int_{a_0}^a \frac{da}{aH}\times\nonumber \label{48}\\
    &&\sqrt{ 1-\frac{\Omega_{D}}{36c^4(1 + \Omega_k)} -\frac{1+\Omega_k }{3\Omega_D}},\\
    \phi\left(a\right)_{Dpl} - \phi\left(a_0\right)_{Dpl}&=&\int_{a_0}^a \frac{da}{aH}\times\nonumber \label{}\\
    &&\sqrt{1-\Omega_{D}\left(\frac{M_p^4 \mathcal{G}}{4\rho_{Dpl}^2}\right)- \frac{1 }{3\Omega_{D}} },\\
    \phi\left(a\right)_{Dlog} - \phi\left(a_0\right)_{Dlog}&=&\int_{a_0}^a \frac{da}{aH}\times\nonumber \label{}\\
    &&\sqrt{1-\Omega_{D}\left(\frac{M_p^4 \mathcal{G}}{4\rho_{Dlog}^2}\right)- \frac{1 }{3\Omega_{D}} },\\
    \phi\left(a\right)_{Dpl,k} - \phi\left(a_0\right)_{Dpl,k}&=&\int_{a_0}^a \frac{da}{aH}\times\nonumber \label{}\\
    &&\sqrt{1-\frac{\Omega_{D}}{1+\Omega_k}\left(\frac{M_p^4 \mathcal{G}}{4\rho_{Dpl}^2}\right)- \frac{1+\Omega_k }{3\Omega_{D}} },\\
    \phi\left(a\right)_{Dlog,k} - \phi\left(a_0\right)_{Dlog,k}&=&\int_{a_0}^a \frac{da}{aH}\times\nonumber \label{}\\
    &&\sqrt{1-\frac{\Omega_{D}}{1+\Omega_k}\left(\frac{M_p^4 \mathcal{G}}{4\rho_{Dlog}^2}\right)- \frac{1+\Omega_k }{3\Omega_{D}} },
\end{eqnarray}
where the quantity $a_0$ indicates the present value of the scale factor. \\
In the limiting case of a flat Dark Dominated Universe, we derive that the scalar field and the potential of the tachyon are given, respectively, by:
\begin{eqnarray}
\dot{\phi}^2_{DD}&=& 1 +\omega_D(t)_{DD}= \frac{2}{3} - \frac{1}{324c^4 M_p^4},\label{49}\\
V(t)_{DD}&=& \rho_{D}(t)_{DD}\sqrt{-\omega_D(t)_{DD}} \nonumber \\
&=& \left[ \frac{1}{\left( 1 - \frac{1}{216 c^4 M_p^4} \right) t + \frac{1}{H_0}}\right]^2\sqrt{\frac{1}{3} + \frac{1}{324c^4 M_p^4}}\label{50}.
\end{eqnarray}
From Eq, (\ref{}) we find:
\begin{eqnarray}
\phi_{DD}(t)&=& \sqrt{ \frac{2}{3} - \frac{1}{324c^4 M_p^4}}\cdot t+constant.\label{49}
\end{eqnarray}

\subsection{k-essence Scalar Field Model}
A class of scalar field theories in which the kinetic contribution enters the Lagrangian in a non-standard (non-canonical) way is commonly referred to as k-essence. This framework was originally inspired by the Born–Infeld action arising in string theory and has been extensively employed to account for the observed late-time acceleration of the Universe \citep{sen-2002,lambert}. In general, the k-essence action can be written in terms of the scalar field $\phi$ and its kinetic energy variable $\chi = \dot{\phi}^2/2$ as
\citep{chiba-07-2000,armendariz-11-2000,armendariz-05-2001}:
\begin{eqnarray}
S=\int d^4x \sqrt{-g}\,p\left(\phi, \chi \right),\label{51}
\end{eqnarray}
where the Lagrangian density $p(\phi,\chi)$ is identified with an effective pressure. Within this framework, the pressure and the corresponding energy density associated with the scalar field $\phi$ can be expressed, respectively, as:
\begin{eqnarray}
    p\left(\phi, \chi \right)&=&f\left(\phi\right)\left( -\chi+\chi ^2   \right), \label{52}\\
        \rho\left(\phi, \chi \right)&=&f\left(\phi\right)\left(-\chi+3\chi ^2\right).\label{53}
\end{eqnarray}
The EoS parameter associated with the k-essence scalar field can therefore be expressed as
\begin{eqnarray}
    \omega _K= \frac{p\left(\phi, \chi \right)}{\rho\left(\phi, \chi \right)}=\frac{\chi-1}{3\chi -1}.\label{54}
\end{eqnarray}
Inspection of Eq. (\ref{54}) shows that the k-essence framework allows for a phantom regime ($\omega_K < -1$) when the kinetic variable $\chi$ satisfies the condition $1/3 < \chi < 1/2$.

To model the interacting HDE component through a k-essence description, we assume a dynamical correspondence between the k-essence EoS parameter $\omega_K$ and the EoS parameter $\omega_D$ of the DE models we consider.

By matching Eq. (\ref{54}) with the different expressions of $\omega_D$ a previously obtained for the DE models we study, the kinetic term $\chi$ can be reconstructed as:
\begin{eqnarray}
    \chi_{D} &=& \frac{w_{D}-1}{3w_{D}-1},\\
      \chi_{D,k} &=& \frac{w_{D,k}-1}{3w_{D,k}-1},\\
      \chi_{Dpl} &=& \frac{w_{Dpl}-1}{3w_{Dpl}-1},\\
      \chi_{Dlog} &=& \frac{w_{Dlog}-1}{3w_{Dlog}-1},\\
      \chi_{Dpl,k} &=& \frac{w_{Dpl,k}-1}{3w_{Dpl,k}-1},\\
      \chi_{Dlog,k} &=& \frac{w_{Dlog,k}-1}{3w_{Dlog,k}-1}.
\end{eqnarray}

Furthermore, by matching Eq. (\ref{53}) with the different expressions of $\omega_D$ previously obtained, one can directly derive the following relations for $f(\phi)$:
\begin{eqnarray}
    f\left(\phi \right)_{D}&=&\frac{\rho_{D}}{\chi_{D}(3\chi_{D}-1)},\label{}\\
    f\left(\phi \right)_{D,k}&=&\frac{\rho_{D,k}}{\chi_{D,k}(3\chi_{D,k}-1)},\label{}\\
    f\left(\phi \right)_{Dpl}&=&\frac{\rho_{Dpl}}{\chi_{Dpl}(3\chi_{Dpl}-1)},\label{}\\
    f\left(\phi \right)_{Dlog}&=&\frac{\rho_{Dlog}}{\chi_{Dlog}(3\chi_{Dlog}-1)},\label{}\\
    f\left(\phi \right)_{Dpl,k}&=&\frac{\rho_{Dpl,k}}{\chi_{Dpl,k}(3\chi_{Dpl,k}-1)},\label{}\\
    f\left(\phi \right)_{Dlog,k}&=&\frac{\rho_{Dlog,k}}{\chi_{Dlog,k}(3\chi_{Dlog,k}-1)}.\label{}
\end{eqnarray}
Considering the relation $\dot{\phi}^2=2\chi$ and $\dot{\phi}=\phi'H$, it is possible to write:
\begin{eqnarray}
 \phi_{D}'= \frac{\sqrt{2}}{H}\sqrt{  \frac{w_{D}-1}{3w_{D}-1}},\label{}\\
 \phi_{D,k}'= \frac{\sqrt{2}}{H}\sqrt{ \frac{w_{D,k}-1}{3w_{D,k}-1} },\label{}\\
 \phi_{Dpl}'= \frac{\sqrt{2}}{H}\sqrt{ \frac{w_{Dpl}-1}{3w_{Dpl}-1} },\label{}\\
  \phi_{Dlog}'= \frac{\sqrt{2}}{H}\sqrt{  \frac{w_{Dlog}-1}{3w_{Dlog}-1}},\label{}\\
 \phi_{Dpl,k}'= \frac{\sqrt{2}}{H}\sqrt{ \frac{w_{Dpl,k}-1}{3w_{Dpl,k}-1} },\label{}\\
 \phi_{Dlog,k}'= \frac{\sqrt{2}}{H}\sqrt{  \frac{w_{Dlog,k}-1}{3w_{Dlog,k}-1}}.\label{}
\end{eqnarray}
Integrating the above relations for $\phi'$, we derive the evolutionary forms of the k-essence scalar field model as follows:
\begin{eqnarray}
    \phi\left(a\right)_{D} -    \phi\left(a_0\right)_{D} &=& \sqrt{2} \int_{a_0}^a \frac{da}{aH}\cdot \sqrt{  \frac{w_{D}-1}{3w_{D}-1}},\label{}\\
     \phi\left(a\right)_{D,k} -    \phi\left(a_0\right)_{D,k} &=& \sqrt{2} \int_{a_0}^a \frac{da}{aH}\cdot \sqrt{  \frac{w_{D,k}-1}{3w_{D,k}-1}} ,\label{}\\
     \phi\left(a\right)_{Dpl} -    \phi\left(a_0\right)_{Dpl} &=& \sqrt{2} \int_{a_0}^a \frac{da}{aH}\cdot \sqrt{  \frac{w_{Dpl}-1}{3w_{Dpl}-1}} ,\label{}\\
     \phi\left(a\right)_{Dlog} -    \phi\left(a_0\right)_{Dlog} &=& \sqrt{2} \int_{a_0}^a \frac{da}{aH}\cdot \sqrt{  \frac{w_{Dlog}-1}{3w_{Dlog}-1}} ,\label{}\\
     \phi\left(a\right)_{Dpl,k} -    \phi\left(a_0\right)_{Dpl,k} &=& \sqrt{2} \int_{a_0}^a \frac{da}{aH}\cdot \sqrt{  \frac{w_{Dpl,k}-1}{3w_{Dpl,k}-1}},\label{}\\
     \phi\left(a\right)_{Dlog,k} -    \phi\left(a_0\right)_{Dlog,k} &=& \sqrt{2} \int_{a_0}^a \frac{da}{aH}\cdot\sqrt{  \frac{w_{Dlog,k}-1}{3w_{Dlog,k}-1}}.\label{}
\end{eqnarray}
In the limiting case of a flat Dark Dominated Universe, the scalar field and potential of k-essence scalar field model reduce, respectively, to:
\begin{eqnarray}
\dot{\phi}^2=2\chi(t)&=&\frac{2\omega_D(t)_{DD}-2}{3\omega_D(t)_{DD}-1}\nonumber \\
&=&\frac{864c^4 M_p^4 + 2}{648c^4 M_p^4 + 3},\label{locat1}\\
f(t)&=&\frac{\rho_{D}(t)_{DD}}{\chi(t)(3\chi(t)-1)}\nonumber \\
&=&\frac{3 \left[ 4(108c^4M_p^4)^2 + 4(108c^4M_p^4) + 1 \right]}{2(108c^4M_p^4) \left[ 4(108c^4M_p^4) + 1 \right]}
 \cdot \left[ \frac{1}{\left( 1 - \frac{1}{216 c^4 M_p^4} \right) t + \frac{1}{H_0}}\right]^2.\label{60}
\end{eqnarray}
Integrating Eq. (\ref{locat1}), we obtain: 
\begin{eqnarray}
\phi(t)
&=&\sqrt{\frac{864c^4 M_p^4 + 2}{648c^4 M_p^4 + 3}} \cdot t + constant.
\end{eqnarray}

\subsection{Quintessence Scalar Field Model}
In the quintessence framework, the late-time acceleration of the universe is accounted for by a minimally coupled, spatially homogeneous scalar field $\phi(t)$ governed by an appropriate potential $V\left(\phi\right)$. The action describing this system is given by \citep{copeland-2006}:
\begin{eqnarray}
    S=\int d^4x \sqrt{-g}\,\left[-\frac{1}{2}g^{\mu \nu} \partial _{\mu} \phi   \partial _{\nu} \phi - V\left( \phi \right)  \right].\label{69}
\end{eqnarray}
By performing a variation of the action in Eq. (\ref{69}) with respect to the metric tensor $g^{\mu \nu}$, the energy-momentum tensor $T_{\mu \nu}$ of the scalar field is obtained as:
\begin{eqnarray}
T_{\mu \nu}=\frac{2}{\sqrt{-g}} \frac{\delta S}{\delta g^{\mu \nu}},\label{70}
\end{eqnarray}
leading to:
\begin{eqnarray}
    T_{\mu \nu}=\partial _{\mu} \phi   \partial _{\nu} \phi - g_{\mu \nu}\left[\frac{1}{2}g^{\alpha \beta} \partial _{\alpha} \phi   \partial _{\beta} \phi + V\left( \phi \right)  \right].\label{71}
\end{eqnarray}
For the quintessence scalar field model, the associated energy density $\rho_Q$ and pressure $p_Q$ are expressed, respectively, by:
\begin{eqnarray}
    \rho_Q&=&-T_0^0=\frac{1}{2}\dot{\phi}^2+V\left(\phi\right),\label{72}\\
    p_Q&=&T_i^i=\frac{1}{2}\dot{\phi}^2-V\left(\phi\right).\label{73}
\end{eqnarray}
while the EoS parameter $\omega_Q$ can be written as:
\begin{eqnarray}    \omega_Q=\frac{p_Q}{\rho_Q}=\frac{\dot{\phi}^2-2V\left(\phi\right)}{\dot{\phi}^2+2V\left(\phi\right)}.\label{74}
\end{eqnarray}

It follows from Eq. (\ref{74}) that cosmic acceleration occurs when $\omega_Q < -1/3$, which corresponds to the regime $\dot{\phi}^2<V\left(\phi\right)$. To establish a clear correspondence between the interacting scenario and the quintessence DE framework, we map the respective equations of state by setting $\omega_Q=\omega_{D}$ and we also match the energy densities via $\rho_Q=\rho_{D}$. This matching procedure yields:
\begin{eqnarray}
    \dot{\phi}^2&=&\left(1+\omega_{D} \right)\rho_{D}, \label{75}\\
    V\left( \phi \right) &=& \frac{1}{2}\left(1-\omega_{D} \right)\rho_{D}.\label{76}
\end{eqnarray}
Inserting the expressions of $\omega_D$ we derived for the different cases into Eqs. (\ref{75}) and (\ref{76}) allows us to extract the analytical forms of both the kinetic term $\dot{\phi}^2$ and the potential energy $V\left( \phi \right)$ as follows:
\begin{eqnarray}
    \dot{\phi}^2_{D}&=&\left(1+\omega_{D} \right)\rho_{D}, \label{}\\
    V\left( \phi \right)_{D} &=& \frac{1}{2}\left(1-\omega_{D} \right)\rho_{D},\label{}\\
      \dot{\phi}^2_{D,k}&=&\left(1+\omega_{D,k} \right)\rho_{D,k}, \label{}\\
    V\left( \phi \right)_{D,k} &=& \frac{1}{2}\left(1-\omega_{D,k} \right)\rho_{D,k},\label{}\\
      \dot{\phi}^2_{Dpl}&=&\left(1+\omega_{Dpl} \right)\rho_{Dpl}, \label{}\\
    V\left( \phi \right)_{Dpl} &=& \frac{1}{2}\left(1-\omega_{Dpl} \right)\rho_{Dpl},\label{}\\
      \dot{\phi}^2_{Dlog}&=&\left(1+\omega_{Dlog} \right)\rho_{Dlog}, \label{}\\
    V\left( \phi \right)_{Dlog} &=& \frac{1}{2}\left(1-\omega_{Dlog} \right)\rho_{Dlog},\label{}\\
      \dot{\phi}^2_{Dpl,k}&=&\left(1+\omega_{Dpl,k} \right)\rho_{Dpl,k}, \label{}\\
    V\left( \phi \right)_{Dpl,k} &=& \frac{1}{2}\left(1-\omega_{Dpl,k} \right)\rho_{Dpl,k},\label{}\\
      \dot{\phi}^2_{Dlog,k}&=&\left(1+\omega_{Dlog,k} \right)\rho_{Dlog,k}, \label{}\\
    V\left( \phi \right)_{Dlog,k} &=& \frac{1}{2}\left(1-\omega_{Dlog,k} \right)\rho_{Dlog,k}.\label{}
\end{eqnarray}
Using the relation $\dot{\phi}=\phi' H$, we derive:
\begin{eqnarray}
\phi\left(a\right)_{D} - \phi \left(a_0\right)_{D}&=& \int_{a_0}^{a}\frac{da}{a}\sqrt{3M_p^2\Omega_{D}}\times \left( 1+\omega_{D}   \right)^{1/2},\label{}\\
\phi\left(a\right)_{D,k} - \phi \left(a_0\right)_{D,k}&=& \int_{a_0}^{a}\frac{da}{a}\sqrt{3M_p^2\Omega_{D}}\times\left(  1+\omega_{D,k}  \right)^{1/2},\label{}\\
\phi\left(a\right)_{Dpl} - \phi \left(a_0\right)_{Dpl}&=& \int_{a_0}^{a}\frac{da}{a}\sqrt{3M_p^2\Omega_{D}}\times \left( 1+\omega_{Dpl}   \right)^{1/2},\label{}\\
\phi\left(a\right)_{Dlog} - \phi \left(a_0\right)_{Dlog}&=& \int_{a_0}^{a}\frac{da}{a}\sqrt{3M_p^2\Omega_{D}}\times \left( 1+\omega_{Dlog}   \right)^{1/2},\label{}\\
\phi\left(a\right)_{Dpl,k} - \phi \left(a_0\right)_{Dpl,k}&=& \int_{a_0}^{a}\frac{da}{a}\sqrt{3M_p^2\Omega_{D}}\times \left(  1+\omega_{Dpl,k}  \right)^{1/2},\label{}\\
\phi\left(a\right)_{Dlog,k} - \phi \left(a_0\right)_{Dlog,k}&=& \int_{a_0}^{a}\frac{da}{a}\sqrt{3M_p^2\Omega_{D}}\times \left( 1+\omega_{Dlog,k}   \right)^{1/2}.\label{}
\end{eqnarray}

In the limiting case of a flat Dark Dominated universe, we can write:
\begin{eqnarray}
\dot \phi^2(t)&=&\left(\frac{2}{3} - \frac{1}{324c^4 M_p^4} \right) \left[ \frac{1}{\left( 1 - \frac{1}{216 c^4 M_p^4} \right) t + \frac{1}{H_0}}\right]^2,\label{80sinner}\\
V(t)&=&\frac{1}{2}\left(\frac{4}{3} + \frac{1}{324c^4 M_p^4} \right) \left[ \frac{1}{\left( 1 - \frac{1}{216 c^4 M_p^4} \right) t + \frac{1}{H_0}}\right]^2\label{81}.
\end{eqnarray}
Integrating Eq. (\ref{80sinner}), we obtain the following result:
\begin{eqnarray}
    \phi(t) = \frac{\sqrt{\frac{2}{3} - \frac{1}{324c^4 M_p^4}}}{1 - \frac{1}{216 c^4 M_p^4}} \ln\left| \left( 1 - \frac{1}{216 c^4 M_p^4} \right) t + \frac{1}{H_0} \right| + constant.
\end{eqnarray}

\subsection{Generalized Chaplygin Gas (GCG) Model}
Next, we focus on the Generalized Chaplygin Gas (GCG) framework. Initially, Kamenshchik \emph{et al.}~\cite{gcg1} introduced the standard Chaplygin Gas (CG) model, a homogeneous scenario characterized by a single fluid governed by the equation of state (EoS) $p = -\frac{A_0}{\rho}$, where $p$ and $\rho$ represent the fluid pressure and energy density, while $A_0$ is a positive constant. This setup was later extended into the GCG model.A remarkably attractive property of the GCG paradigm is its capacity to smoothly interpolate from an early dust-dominated era to a late-time phase of cosmic acceleration. Consequently, this model offers a highly competitive fit when confronted with modern observational data~\cite{gcg6}.\\
The EoS behavior for the GCG framework is analytically expressed as~\cite{gcg9,gcg16,gcg17}:
\begin{eqnarray}
    p_D = -\frac{D}{\rho_D^{\theta}}, \label{gcg1}
\end{eqnarray}
Here, $D$ and $\theta$ represent two constant parameters, where $D > 0$ and $\theta$ is restricted to the interval $0 < \theta < 1$. It is worth noting that the standard Chaplygin Gas (CG) model is readily recovered by setting $\theta = 1$. Historically, the EoS in Eq.~(\ref{gcg1}) with $\theta = 1$ was originally introduced in 1904 by Chaplygin within the framework of adiabatic processes~\cite{gcg1}, whereas the generalized scenario involving $\theta \neq 1$ was later investigated in~\cite{bento}. The conceptual framework utilizing the Chaplygin gas to achieve a unified description of DE and DM was first advanced for the case $\theta = 1$ in~\cite{gcg4,gcg5}, and subsequently extended to arbitrary values of $\theta \neq 1$ in~\cite{bento}.

As demonstrated by Gorini \emph{et al.}~\cite{gcg15}, the matter power spectrum predicted within the GCG framework aligns with observational data only under the stringent condition $\theta < 10^{-5}$. Such a tight constraint effectively renders the GCG model observationally indistinguishable from the standard $\Lambda$CDM paradigm. Concurrently, the implications of Chaplygin inflation have been investigated through the lens of Loop Quantum Cosmology in~\cite{gcg7}.

The cosmological evolution of the energy density $\rho_D$ within the GCG framework is governed by:
\begin{eqnarray}
    \rho_D = \left[D + \frac{B}{a^{3\left(\theta+1\right)}}\right]^{\frac{1}{\theta+1}}, \label{gcg2}
\end{eqnarray}
where the quantity $B$ represents an integration constant.

In principle, the parameter $\theta$ in Eq.~(\ref{gcg2}) can take any positive value; however, to preserve causality, the speed of sound $c_s^2 = \frac{D\theta}{\rho^{\theta+1}}$ must be strictly subluminal. Moreover, to guarantee a well-behaved evolution for the energy density fluctuations, one must restrict the analysis to the range $0 < \theta < 1$~\cite{bento}. Next, we establish the correspondence needed to reconstruct the potential and dynamics of this scalar field description within our current DE scenario. The respective expressions for the energy density $\rho_D$ and pressure $p_D$ of the homogeneous, time-dependent scalar field $\phi$ read:
\begin{eqnarray}
\rho_D &=& \frac{1}{2}\dot{\phi}^2 + V\left(\phi\right), \label{gcg3}\\
p_D &=& \frac{1}{2}\dot{\phi}^2 - V\left(\phi\right). \label{gcg4}
\end{eqnarray}
Utilizing Eqs.~(\ref{gcg3}) and (\ref{gcg4}), the EoS parameter $\omega_D$ for the GCG model can be explicitly written as:
\begin{eqnarray}
\omega_D &=& \frac{\frac{1}{2} \dot{\phi}^2 - V\left(\phi\right)}{\frac{1}{2} \dot{\phi}^2 + V\left(\phi\right)} = \frac{\dot{\phi}^2 - 2 V\left(\phi\right)}{\dot{\phi}^2 + 2 V\left(\phi\right)}. \label{gcg5}
\end{eqnarray}

The combination of Eqs.~(\ref{gcg3}) and (\ref{gcg4}) readily provides the expression for the kinetic energy term $\dot{\phi}^2$, namely
\begin{eqnarray}
    \dot{\phi}^2 &=& \rho_D + p_D. \label{gcg6-1}
\end{eqnarray}
Inserting the GCG pressure relation from Eq.~(\ref{gcg1}) into Eq.~(\ref{gcg6-1}), we find
\begin{eqnarray}
    \dot{\phi}^2 &=& \rho_D - \frac{D}{\rho_D^\theta} = \rho_D \left(1 - \frac{D}{\rho_D^{\theta+1}}\right). \label{gcg6-11}
\end{eqnarray}

By taking into account the kinematic relation $\dot{\phi} = H \phi'$ and incorporating the explicit form of $\rho_D$ from Eq.~(\ref{gcg2}), the dynamical evolution of the scalar field $\phi$ can be recast as
\begin{eqnarray}
    \phi &=& \int_{a_0}^a \sqrt{3\Omega_D} \sqrt{1 - \frac{D}{D + \frac{B}{a^{3(\theta+1)}}}} \frac{da}{a}, \label{gcg6-1newint}
\end{eqnarray}
whose solution, in the limiting case of a flat Dark Dominated Universe, reads
\begin{eqnarray}
 \phi(a) &=& \frac{2 \sqrt{3}}{3(1+\theta)} \times \nonumber \\
 && \left[\log\left(a^{\frac{3}{2}(1+\theta)}\right) - \log\left(B + \sqrt{B\left(B + a^{3(1+\theta)} D\right)}\right)\right]. \label{murano4}
\end{eqnarray}

By subtracting Eq.~(\ref{gcg4}) from Eq.~(\ref{gcg3}) and incorporating the pressure definition from Eq.~(\ref{gcg1}), the scalar potential $V(\phi)$ is found to be
\begin{eqnarray}
 V(\phi) &=& \frac{1}{2} (\rho_D - p_D) = \frac{\rho_D}{2} \left(1 + \frac{D}{\rho_D^{\theta+1}}\right). \label{gcg7-1}
\end{eqnarray}
Then, substituting the general solution for $\rho_D$ from Eq.~(\ref{gcg2}) allows us to express $V(\phi)$ in its explicit form:
\begin{eqnarray}
V(\phi) &=& \frac{1}{2}\left[D + \frac{B}{a^{3(\theta+1)}}\right]^{\frac{1}{\theta+1}} + \frac{1}{2} \frac{D}{\left[D + \frac{B}{a^{3(\theta+1)}}\right]^{\frac{\theta}{\theta+1}}}. \label{gcg7-1new}
\end{eqnarray}

We now proceed to isolate the parameters $D$ and $B$ in terms of background cosmological quantities. \\
Dividing both sides of Eq.~(\ref{gcg1}) by $\rho_D$ allows us to express the EoS parameter $\omega_D$ as:
\begin{eqnarray}
    \omega_D = - \frac{D}{\rho_D^{\theta+1}}, \label{gcg8}
\end{eqnarray}
yielding the following expression for the parameter $D$:
\begin{eqnarray}
    D = - \omega_D \rho_D^{\theta+1}. \label{gcg9}
\end{eqnarray}
From Eq.~(\ref{gcg2}), we obtain that the integration constant $B$ is given by the following general relation
\begin{eqnarray}
B = a^{3(\theta+1)} \left(\rho_D^{\theta+1} - D \right), \label{murano5}
\end{eqnarray}
Inserting Eq.~(\ref{gcg9}) into this relation to eliminate $D$, we find:
\begin{eqnarray}
B = \left(a^3 \rho_D\right)^{\theta+1} (1 + \omega_D). \label{BBB}
\end{eqnarray}
For the models we studied in this paper, we obtain the following results for the parameters $D$ and $B$:
\begin{eqnarray}
D_{D}&=& -\rho_{D}^{\theta + 1}\cdot \omega_{D}, \label{}\\
B_{D}&=&\left(a^3\rho_{D,k}\right)^{\theta+1}\cdot \left( 1+\omega_{D,k}\right),\label{} \\
D_{D,k}&=& -\rho_{D,k}^{\theta + 1}\cdot \omega_{D,k}, \label{}\\
B_{D,k}&=&\left(a^3\rho_{D,k}\right)^{\theta+1}\cdot \left( 1+\omega_{D,k}\right),\label{} \\
D_{Dpl}&=& -\rho_{Dpl}^{\theta + 1}\cdot \omega_{Dpl}, \label{}\\
B_{Dpl}&=&\left(a^3\rho_{Dpl}\right)^{\theta+1}\cdot \left( 1+\omega_{Dpl}\right),\label{} \\
D_{Dlog}&=& -\rho_{Dlog}^{\theta + 1}\cdot \omega_{Dlog}, \label{}\\
B_{Dlog}&=&\left(a^3\rho_{Dlog}\right)^{\theta+1}\cdot \left( 1+\omega_{Dlog}\right),\label{} \\
D_{Dpl,k}&=& -\rho_{Dpl,k}^{\theta + 1}\cdot \omega_{Dpl,k}, \label{}\\
B_{Dpl,k}&=&\left(a^3\rho_{Dpl,k}\right)^{\theta+1}\cdot \left( 1+\omega_{D}\right),\label{} \\
D_{Dlog,k}&=& -\rho_{Dlog,k}^{\theta + 1}\cdot \omega_{Dlog,k}, \label{}\\
B_{Dlog,k}&=&\left(a^3\rho_{Dlog,k}\right)^{\theta+1}\cdot \left( 1+\omega_{Dlog,k}\right).\label{} 
\end{eqnarray}

In the limiting case of a flat Dark Dominated Universe, we find that the expressions of $D_{ Dark}$ and $B_{ Dark}$ are given by the following two relations:
\begin{eqnarray}
D_{ Dark} &=& \left(\frac{1}{3} + \frac{1}{324c^4 M_p^4} \right)\cdot\left[ \frac{1}{\left( 1 - \frac{1}{216 c^4 M_p^4} \right) t + \frac{1}{H_0}}\right]^{2(\theta +1)}
, \label{murano10}\\
B_{ Dark} &=& \left\{a_0^3 H_0^2 \left[ 1 + \left( 1 - \frac{1}{216 c^4 M_p^4} \right) H_0 t \right]^{\frac{3}{1 - \frac{1}{216 c^4 M_p^4}} - 2}   \right\}^{\theta+1}\times \nonumber \\
&&\left(1 -\frac{1}{3} - \frac{1}{324c^4 M_p^4}\right). \label{murano11}
\end{eqnarray}

\subsection{Yang-Mills (YM) Model}

We now turn our attention to the Yang-Mills (YM) framework. In recent years, the YM field has been widely investigated as a compelling candidate to explain the nature of DE~\cite{ym1,ym9-3,ym9-5,ym9-7}. Two primary arguments motivate the choice of a YM field over standard scalar field prototypes. First, its physical foundations are more rigorously grounded within fundamental particle physics. Second, the YM framework naturally allows for violations of the weak energy condition, thereby opening up novel and rich cosmological scenarios. The specific YM setup analyzed here displays several remarkable features. Because it serves as a cornerstone of gauge theories—where interactions are mediated by gauge bosons - it fits seamlessly into grand unified theories. Furthermore, the EoS parameter characterizing the effective Yang-Mills Condensate (YMC) exhibits a distinctly different behavior compared to conventional matter or standard scalar fields. Remarkably, the YMC can smoothly access both the quintessence regime ($-1 < \omega < 0$) and the phantom domain ($\omega < -1$).

Within the effective YMC DE framework, the effective Yang-Mills field Lagrangian density $L_{\mathrm{YMC}}$ is defined as
\begin{eqnarray}
L_{YMC} = \frac{bF}{2} \left( \ln \left| \frac{F}{\kappa^2} \right| - 1 \right), \label{murano117}
\end{eqnarray}
where $\kappa$ denotes the renormalization scale carrying the dimension of mass squared, while $F$ serves as the order parameter of the YMC. The quantity $F$ is explicitly given by
\begin{eqnarray}
F = -\frac{1}{2} F_{\mu \nu}^\alpha F^{\alpha \mu \nu} = E^2 - B^2, \label{murano118}
\end{eqnarray}
where the quantity $F_{\mu \nu}^\alpha$ represents the Yang-Mills field strength tensor, with $E$ and $B$ indicating the effective electric and magnetic field components, respectively. In the purely electric configuration, one sets $B = 0$, which directly simplifies the order parameter to $F = E^2$.

Here, the parameter $b$ is the Callan-Symanzik beta function coefficient~\cite{ym18,ym18-1}, which reads
\begin{equation}
b = \frac{11N - 2N_f}{24\pi^2} \label{murano119}
\end{equation}
for an $SU(N)$ gauge symmetry, with $N_f$ representing the flavor number. In the $SU(2)$ sector, this parameter takes the value $b = \frac{11}{12\pi^2}$ in the pure gauge limit, and shifts to $b = \frac{5}{12\pi^2}$ for $N_f = 6$. Alternatively, when dealing with the $SU(3)$ group, the effective Lagrangian formulation offers a suitable phenomenological description of asymptotic freedom for hadronic quarks~\cite{ym21,ym21-1}.

It is crucial to emphasize that the $SU(2)$ Yang-Mills field introduced herein as a DE candidate must not be identified with the QCD gluon fields or the electroweak gauge bosons ($W^{\pm}$ and $Z^0$). Indeed, the energy scale of the Yang-Mills condensate (YMC) is governed by the parameter $\kappa^{1/2} \approx 10^{-3} \text{ eV}$, a value many orders of magnitude below the characteristic scales of QCD and electroweak interactions.Physically, the effective Lagrangian formulation presented in Eq.~\eqref{murano117} stems from a 1-loop quantum correction framework~\cite{ym21,ym21-1}. The standard classical $SU(N)$ Yang-Mills Lagrangian reads
\begin{equation}
L = \frac{1}{2g_0^2} F, \label{murano120}
\end{equation}where $g_0$ represents the bare coupling constant. Incorporating 1-loop quantum corrections induces a running behavior for the gauge coupling $g$, which replaces the bare parameter according to
\begin{equation}
g_0^2 \rightarrow g^2 = \frac{48\pi^2}{11N \ln\left(\frac{k^2}{k_0^2}\right)} = \frac{2}{b \ln\left(\frac{k^2}{k_0^2}\right)}, \label{murano121}
\end{equation}
where the quantity $k^2$ is the momentum transfer and $k_0^2$ denotes the reference renormalization scale.Within the effective field theory approach, the momentum scale $k^2$ is mapped onto the field strength order parameter $F$ via the standard prescription
\begin{equation}
\ln\left(\frac{k^2}{k_0^2}\right) \rightarrow \ln \left| \frac{F}{\kappa^2 e} \right| = \ln \left| \frac{F}{\kappa^2} \right| - 1. \label{murano122}
\end{equation}
By substituting Eq.~\eqref{murano122} into Eq.~\eqref{murano120} alongside the running coupling relation, one straightforwardly recovers the effective Lagrangian density of Eq.~\eqref{murano117}.

The effective YM action considered here exhibits several compelling characteristics, such as Lorentz and gauge invariance, asymptotic freedom, and compliance with the expected trace anomaly. Due to the logarithmic structure involving the field strength invariant, this Lagrangian shares a clear mathematical analogy with the Coleman-Weinberg scalar effective potential~\cite{ym19} and the Parker-Raval gravity model~\cite{ym20}. Crucially, the renormalization scale $\kappa$ constitutes the only free parameter of the theory. In sharp contrast to scalar field models of DE, the functional structure of the YM Lagrangian cannot be adjusted arbitrarily, as it is uniquely determined by 1-loop quantum corrections. Utilizing the effective Lagrangian from Eq.~\eqref{murano117}, the energy density $\rho_y$ and pressure $p_y$ associated with the YMC can be cast as:
\begin{eqnarray}
\rho_y &=& \frac{\epsilon E^2}{2} + \frac{b E^2}{2}, \label{murano123}\\
p_y &=& \frac{\epsilon E^2}{6} - \frac{b E^2}{2}, \label{murano124}
\end{eqnarray}
where the quantity \(\epsilon\) represents the dielectric constant of the YMC, which is generally defined in the following way:
\begin{eqnarray}
\epsilon = 2 \frac{\partial L_{eff}}{\partial F} = b \ln \left| \frac{F}{\kappa^2} \right|. \label{murano125}
\end{eqnarray}

Eqs. \eqref{murano123} and \eqref{murano124} can be also written as
\begin{eqnarray}
\rho_y &=& \frac{1}{2} b \kappa^2 (y+1) e^{y}, \label{murano126}\\
p_y &=& \frac{1}{6} b \kappa^2 (y-3) e^{y}, \label{murano127}
\end{eqnarray}
or equivalently:
\begin{eqnarray}
\rho_y &=& \frac{1}{2} (y+1) b E^2, \label{murano128} \\
p_y &=& \frac{1}{6} (y-3) b E^2, \label{murano129}
\end{eqnarray}
where it was introduced the dimensionless parameter $y$, given by:
\begin{eqnarray}
y = \frac{\epsilon}{b} = \ln \left| \frac{F}{\kappa^2} \right| = \ln \left| \frac{E^2}{\kappa^2} \right|. \label{defiy}
\end{eqnarray}

By utilizing the expressions for $\rho_y$ and $p_y$ given in Eqs.~\eqref{murano126} and \eqref{murano127} (or, equivalently, Eqs.~\eqref{murano128} and \eqref{murano129}), the equation of state (EoS) parameter $\omega_y$ of the YMC framework can be written as:
\begin{eqnarray}
\omega_y = \frac{p_y}{\rho_y} = \frac{y - 3}{3(y + 1)}. \label{omegay}
\end{eqnarray}

At the critical point where $\epsilon = 0$ (corresponding to $y = 0$), the EoS parameter reduces to $\omega_y = -1$, thereby driving a de Sitter expansion of the Universe. In the vicinity of this configuration, the condition $\epsilon < 0$ leads to the phantom regime $\omega_y < -1$, whereas $\epsilon > 0$ yields $\omega_y > -1$. Consequently, as emphasized above, the YMC framework naturally accommodates both the quintessence domain ($-1 < \omega_y < 0$) and the phantom-like behavior ($\omega_y < -1$).

By inverting Eq.~\eqref{omegay}, we can express the dimensionless variable $y$ in terms of the EoS parameter as
\begin{equation}
y = - \frac{3(\omega_y + 1)}{3\omega_y - 1}. \label{murano130}
\end{equation}
Requiring the energy density $\rho_y$ to remain strictly positive for physical consistency mandates that $y > 1$. This restriction implies that the order parameter $F$ must obey the threshold
\begin{equation}
F > \frac{\kappa^2}{e} \approx 0.368, \kappa^2.
\end{equation}
Prior to investigating a particular cosmological background, it is useful to study the dependence of $\omega_y$ on the condensate strength $F$. Notably, the YMC framework features a radiation-like equation of state with
\begin{eqnarray}
p_y = \frac{1}{2} \rho_y
\end{eqnarray}
and the corresponding EoS parameter $\omega_y$
\begin{eqnarray}
\omega_y = \frac{1}{2}
\end{eqnarray}
for large values of the dielectric constant $\epsilon$, i.e. \(\epsilon \gg b\) (implying \(F \gg \kappa^2\)).

Alternatively, when the system crosses the critical point $\epsilon = 0$ (where $F = \kappa^2$), the YMC satisfies a de Sitter equation of state, namely
\begin{align}
\omega_y &= -1, \nonumber \
p_y &= -\rho_y.
\end{align}
This specific setup fixes the YMC energy density to the constant baseline
\begin{equation}
\rho_y = \frac{1}{2} b \kappa^2,
\end{equation}
which serves as the critical energy density governing the acceleration~\cite{ym1}.

The capacity of the YMC equation of state to interpolate smoothly between a radiation character $\omega_y = 1/3$ in the high-energy limit ($F \gg \kappa^2$) and a cosmological constant behavior $\omega_y = -1$ at low energies ($F = \kappa^2$) underpins the realization of cosmological scaling solutions~\cite{ym10-1}. This evolutionary path is well-defined since $\omega_y$ depends continuously on the underlying variable $y$ .Let us now evaluate the conditions required to cross the phantom divide $\omega_y = -1$. Structurally, Eq.~\eqref{omegay} implies that a phantom-like phase ($\omega_y < -1$) can be achieved provided that
\begin{equation}
F < \kappa^2,
\end{equation}
a process that is mathematically smooth with respect to the field invariant $F$. Nevertheless, in a realistic cosmological model where the YMC acts as DE in the presence of matter and radiation, $F$ becomes a dynamical degree of freedom governed by the cosmic time $t$.If the YMC energy density is conserved separately, $\omega_y$ merely targets $-1$ asymptotically. On the other hand, if an explicit interaction allows the YMC to decay into background fluids, the system can cross into the phantom region, asymptotically approaching a steady state with $\omega_y \approx -1.17$ depending on the coupling parameters. Crucially, within this lower sector ($\omega_y < -1$), the physical quantities $\rho_y$, $p_y$, and $\omega_y$ evolve smoothly without encountering any finite-time cosmic singularities, representing a significant conceptual advantage over typical phantom scalar field alternatives.

By equating the YMC equation of state $\omega_y$ with the effective EoS parameter of the specific cosmological model under consideration, the dimensionless variable $y$ can be recast in the following form:
\begin{eqnarray}
y = - \frac{3\left(\omega_D+1\right)}{3\omega_D-1}.\label{murano131}
\end{eqnarray}
which, upon employing the results obtained above, can be written as:
\begin{eqnarray}
y_{D} &=& - \frac{3\left(\omega_{D}+1\right)}{3\omega_{D}-1} ,\label{}\\
y_{D,k} &=& - \frac{3\left(\omega_{D,k}+1\right)}{3\omega_{D,k}-1}, \label{}\\
y_{Dpl} &=& - \frac{3\left(\omega_{Dpl}+1\right)}{3\omega_{Dpl}-1}, \label{}\\
y_{Dlog} &=& - \frac{3\left(\omega_{Dlog}+1\right)}{3\omega_{Dlog}-1}, \label{}\\
y_{Dpl,k} &=& - \frac{3\left(\omega_{Dpl,k}+1\right)}{3\omega_{Dpl,k}-1}, \label{}\\
y_{Dlog,k} &=& - \frac{3\left(\omega_{Dlog,k}+1\right)}{3\omega_{Dlog,k}-1}. \label{}
\end{eqnarray}

In the limiting case of a flat Dark Dominated Universe, we can write:
\begin{eqnarray}
y(t)_{ Dark} &=& -\frac{3\left(\omega_D(t)_{DD}+1\right)}{3\omega_D(t)_{DD}-1}\nonumber \\
&=&\frac{216c^4 M_p^4 - 1}{216c^4 M_p^4 + 1}.\label{murano134}
\end{eqnarray}
From Eq. (\ref{murano134}) we derive that $y$ is never greater than one.

\subsection{Non Linear Electro-Dynamics (NLED) Model}

We now turn our attention to the final framework chosen for this study: the Non-Linear Electro-Dynamics (NLED) model. In recent years, non-linear extensions of Maxwell's electromagnetic theory have attracted significant interest as a novel approach to avoid primordial cosmic singularities. Indeed, exact solutions of the Einstein field equations coupled with NLED source terms demonstrate that non-linear electromagnetic effects become non-negligible and physically acceptable within strong gravitational and magnetic regimes. Furthermore, General Relativity (GR) incorporating NLED corrections offers a viable mechanism to drive the primordial inflationary phase of the Universe. The standard Maxwell Lagrangian density $L_M$ for free electromagnetic fields can be written as~\cite{ele1,ele2}:
\begin{equation}
L_M = - \frac{F^{\mu \nu}F_{\mu \nu}}{4\mu}, \label{murano135}
\end{equation}
where the quantity $F^{\mu \nu}$ denotes the electromagnetic field strength tensor and $\mu$ represents the magnetic permeability of the medium.

We now consider the non-linear generalization of the Maxwell electromagnetic Lagrangian up to second-order invariants of the field strengths, expressed as:
\begin{equation}
L = -\frac{F}{4\mu_0} + \omega F^2 + \eta F^{*2}, \label{murano136}
\end{equation}
where $\omega$ and $\eta$ are two arbitrary coupling constants. Here, the secondary invariant $F^*$ is defined through the relation
\begin{equation}
F^* = F_{\mu \nu}^* F^{\mu \nu}, \label{murano137}
\end{equation}
with $F_{\mu \nu}^*$ denoting the dual electromagnetic field strength tensor.

In what follows, we examine the regime where the homogeneous plasma electric field $E$ rapidly dissipates via charged particle currents, allowing the magnetic field to dominate completely. Setting $E^2 \approx 0$, the first electromagnetic invariant simplifies to
\begin{equation}
F = 2B^2,
\end{equation}
showing that $F$ is uniquely determined by the magnetic field $B$. The resulting thermodynamic pressure $p_{\mathrm{NLED}}$ and energy density $\rho_{\mathrm{NLED}}$ of the NLED component take the forms:
\begin{align}
p_{NLED} &= \frac{B^2}{6\mu}\left(1 - 40 \mu \omega B^2 \right), \label{murano138}\\
\rho_{NLED} &= \frac{B^2}{2\mu}\left(1 - 8 \mu \omega B^2 \right). \label{murano139}
\end{align}

The Weak Energy Condition (WEC), i.e. \(\rho_{NLED} > 0\), is satisfied when
\begin{eqnarray}
B < \frac{1}{2\sqrt{2 \mu \omega}},\label{sinner}
\end{eqnarray}
while the pressure \(p_{NLED}\) is negative defined when
\begin{eqnarray}
B > \frac{1}{2\sqrt{10 \mu \omega}}.
\end{eqnarray}

The magnetic field can successfully drive a DE behavior provided that the Strong Energy Condition (SEC) is violated, i.e., if $\rho_{\mathrm{NLED}} + 3 p_{\mathrm{NLED}} < 0$, which explicitly occurs when
\begin{eqnarray}
B > \frac{1}{2\sqrt{6 \mu \omega}}.
\end{eqnarray}

The EoS parameter \(\omega_{NLED}\) for the Nonlinear Electrodynamics Field (NLED) field is defined as follows
\begin{eqnarray}
\omega_{NLED} = \frac{p_{NLED}}{\rho_{NLED}} = \frac{1 - 40 \mu \omega B^2}{3 \left(1 - 8 \mu \omega B^2\right)}, \label{murano140}
\end{eqnarray}
which can be solved to obtain the following explicit expression for $B^2$:
\begin{eqnarray}
B^2 = \frac{1 - 3 \omega_{NLED}}{8 \mu \omega \left(5 - 3 \omega_{NLED}\right)}. \label{murano141}
\end{eqnarray}

By establishing a direct correspondence between the EoS parameter of the NLED framework and that of the DE density, we obtain:
\begin{eqnarray}
B^2 = \frac{1 - 3 \omega_D}{8 \mu \omega \left(5 - 3 \omega_D\right)}. \label{murano142}
\end{eqnarray}

For the various cosmological frameworks analyzed in this work, the corresponding analytical expressions for the squared magnetic field strength $B^2$ are found to be:
\begin{eqnarray}
B^2_{D} &=& \frac{1-3\omega_{D}}{8\mu \omega\left(5 -3\omega_{D}   \right) }, \label{}\\
B^2_{D,k} &=& \frac{1-3\omega_{D,k}}{8\mu \omega\left(5 -3\omega_{D,k}   \right) }, \label{}\\
B^2_{Dpl} &=& \frac{1-3\omega_{Dpl}}{8\mu \omega\left(5 -3\omega_{Dpl}   \right) } ,\label{}\\
B^2_{Dlog} &=& \frac{1-3\omega_{Dlog}}{8\mu \omega\left(5 -3\omega_{Dlog}   \right) }, \label{}\\
B^2_{Dpl,k} &=& \frac{1-3\omega_{Dpl,k}}{8\mu \omega\left(5 -3\omega_{Dpl,k}   \right) } ,\label{}\\
B^2_{Dlog,k} &=& \frac{1-3\omega_{Dlog,k}}{8\mu \omega\left(5 -3\omega_{Dlog,k}   \right) }. \label{}
\end{eqnarray}

Moreover, in the limiting case of a flat Dark Dominated Universe, we obtain the following result for $B^2(t)$:
\begin{eqnarray}
B^2(t)_{Dark}  &=&\frac{1 - 3 \omega(t)_{DD}}{8 \mu \omega \left(5 - 3 \omega(t)_{DD}\right)} \nonumber \\
&=& \frac{216c^4 M_p^4 + 1}{8 \mu \omega \left(648c^4 M_p^4 + 1\right)}.\label{murano145}
\end{eqnarray}

Therefore, we find that the Weak Energy Condition given in Eq. (\ref{sinner}) is always satisfied.

\section{Conclusions}
In this paper, we considered the Holographic Dark Energy HDE model along with the Power Law and the Logarithmic entropy-corrected versions of the
 HDE model with IR cut-off equivalent to $L= \mathcal{G}^{-1/4}$, where $\mathcal{G}$ indicates the Gauss-Bonnet invariant. The HDE scenario represents an important approach to understanding the nature of DE within the context of quantum gravity. The inclusion of additional correction terms in the HDE energy density is mainly inspired by Loop Quantum Gravity (LQG), which is regarded as one of the leading candidates for a consistent quantum theory of gravity.\\
In this work, we derived the Equation of State (EoS) parameter, the deceleration parameter $q$ and the evolution of the DE fractional density parameter for the considered DE models for four difference scenarios:
\begin{enumerate}
    \item flat non-interacting Universe,
    \item non-flat non-interacting Universe,
    \item flat interacting Universe,
    \item non-flat interacting Universe.
\end{enumerate} 
Furthermore, we investigated the corresponding limiting behavior in the Dark Energy dominated epoch of the Universe. \\
Furthermore, we constructed correspondences between the HDE models considered in this work and several scalar field descriptions of DE, namely the tachyon, k-essence, quintessence, Generalized Chaplygin Gas (GCG), Yang–Mills (YM), and Nonlinear Electrodynamics (NLED) models. Establishing these correspondences provides valuable insight into the mutual relationships between different DE candidates. We also studied these correspondence in the limiting case of a flat Dark Energy dominated Universe.


\begin{thebibliography}{}

\bibitem{perlmutter} Perlmutter, S., et al., Astrophys. J. \textbf{517}, 565 (1999).
\bibitem{astier} Astier, P., et al., Astron. Astrophys. \textbf{447}, 31 (2006).
\bibitem{bennett-09-2003} Bennett, C. L., et al., Astrophys. J. Suppl. \textbf{148}, 1 (2003).
\bibitem{spergel-09-2003} Spergel, D. N., et al., Astrophys. J. Suppl. \textbf{148}, 175 (2003).
\bibitem{tegmark} Tegmark, M., et al., Phys. Rev. D \textbf{69}, 103501 (2004).
\bibitem{abz1} Abazajian, K., et al., Astron. J. \textbf{128}, 502 (2004).
\bibitem{abz2} Abazajian, K., et al., Astron. J. \textbf{129}, 1755 (2005).
\bibitem{allen} Allen, S. W., et al., Mon. Not. Roy. Astron. Soc. \textbf{353}, 457 (2004).
\bibitem{peiris} Peiris, H. V., et al., Astrophys. J. Suppl. \textbf{148}, 213 (2003).
\bibitem{Padmanabhan-07-2003} Padmanabhan, T., Phys. Rept. \textbf{380}, 235 (2003).
\bibitem{peebles} Peebles, P. J. E., \& Ratra, B., Rev. Mod. Phys. \textbf{75}, 559 (2003).
\bibitem{copeland-2006} Copeland, E. J., Sami, M., \& Tsujikawa, S., Int. J. Mod. Phys. D \textbf{15}, 1753 (2006).
\bibitem{weinberg} Weinberg, S., Rev. Mod. Phys. \textbf{61}, 1 (1989).
\bibitem{wetterich} Wetterich, C., Nucl. Phys. B \textbf{302}, 668 (1988).
\bibitem{ratra} Ratra, B., \& Peebles, P. J. E., Phys. Rev. D \textbf{37}, 3406 (1988).
\bibitem{chiba-07-2000} Chiba, T., Okamura, T., \& Yamaguchi, M., Phys. Rev. D \textbf{62}, 023511 (2000).
\bibitem{armendariz-11-2000} Armendariz-Picon, C., Mukhanov, V. F., \& Steinhardt, P. J., Phys. Rev. Lett. \textbf{85}, 4490 (2000).
\bibitem{armendariz-05-2001} Armendariz-Picon, C., Mukhanov, V. F., \& Steinhardt, P. J., Phys. Rev. D \textbf{63}, 103510 (2001).
\bibitem{caldwell-10-2002} Caldwell, R. R., Phys. Lett. B \textbf{545}, 23 (2002).
\bibitem{nojiri-06-2003} Nojiri, S., \& Odintsov, S. D., Phys. Lett. B \textbf{562}, 147 (2003).
\bibitem{nojiri-07-2003} Nojiri, S., \& Odintsov, S. D., Phys. Rev. D \textbf{68}, 123512 (2003).
\bibitem{sen-04-2002} Sen, A., JHEP \textbf{0204}, 048 (2002).
\bibitem{Padmanabhan-06-2002} Padmanabhan, T., Phys. Rev. D \textbf{66}, 021301 (2002).
\bibitem{Padmanabhan-10-2002} Padmanabhan, T., \& Choudhury, T. R., Phys. Rev. D \textbf{66}, 081301 (2002).
\bibitem{gasperini} Gasperini, M., Piazza, F., \& Veneziano, G., Phys. Rev. D \textbf{65}, 023508 (2002).
\bibitem{piazza} Piazza, F., \& Veneziano, G., JCAP \textbf{0407}, 007 (2004).
\bibitem{arkani} Arkani-Hamed, N., Cheng, H. C., Luty, M. A., \& Mukohyama, S., JHEP \textbf{0405}, 074 (2004).
\bibitem{elizalde-08-2004} Elizalde, E., Nojiri, S., \& Odintsov, S. D., Phys. Rev. D \textbf{70}, 043539 (2004).
\bibitem{nojiri-03-2005} Nojiri, S., Odintsov, S. D., \& Tsujikawa, S., Phys. Rev. D \textbf{71}, 063004 (2005).
\bibitem{anisimov} Anisimov, A., Babichev, E., \& Vikman, A., JCAP \textbf{0506}, 006 (2005).
\bibitem{Kamenshchik} Kamenshchik, A. Y., Moschella, U., \& Pasquier, V., Phys. Lett. B \textbf{511}, 265 (2001).
\bibitem{setare-11-2007} Setare, M. R., Phys. Lett. B \textbf{654}, 1 (2007).
\bibitem{deffayet} Deffayet, C., Dvali, G. R., \& Gabadadze, G., Phys. Rev. D \textbf{65}, 044023 (2002).
\bibitem{sahni} Sahni, V., \& Shtanov, Y., JCAP \textbf{0311}, 014 (2003).
\bibitem{cohen} Cohen, A. G., Kaplan, D. B., \& Nelson, A. E., Phys. Rev. Lett. \textbf{82}, 4971 (1999).
\bibitem{horava-08-2000} Horava, P., \& Minic, D., Phys. Rev. Lett. \textbf{85}, 1610 (2000).
\bibitem{setare-01-2007} Setare, M. R., Phys. Lett. B \textbf{644}, 99 (2007).
\bibitem{setare-10-2007} Setare, M. R., Phys. Lett. B \textbf{653}, 145 (2007).
\bibitem{setare-11-2006} Setare, M. R., Phys. Lett. B \textbf{642}, 1 (2006).
\bibitem{setare-05-2007} Setare, M. R., Phys. Lett. B \textbf{648}, 329 (2007).
\bibitem{setare-01-05-2007} Setare, M. R., JCAP \textbf{0701}, 023 (2007).
\bibitem{setare-09-2007} Setare, M. R., Eur. Phys. J. C \textbf{52}, 689 (2007).
\bibitem{cai-12-2007} Cai, R. G., Phys. Lett. B \textbf{657}, 228 (2007).
\bibitem{wei-02-2008} Wei, H., \& Cai, R. G., Phys. Lett. B \textbf{660}, 113 (2008).
\bibitem{miao} Li, M., Li, X. D., Wang, S., \& Wang, Y., Commun. Theor. Phys. \textbf{56}, 525 (2011).
\bibitem{thooft} 't Hooft, G., gr-qc/9310026; Susskind, L., J. Math. Phys. \textbf{36}, 6377 (1995).
\bibitem{fischler} Fischler, W., \& Susskind, L., hep-th/9806039.
\bibitem{huang-08-2004} Huang, Q. G., \& Li, M., JCAP \textbf{0408}, 013 (2004).
\bibitem{hsu} Hsu, S. D. H., Phys. Lett. B \textbf{594}, 13 (2004).
\bibitem{wang-09-2005} Wang, B., Gong, Y. G., \& Abdalla, E., Phys. Lett. B \textbf{624}, 141 (2005).
\bibitem{guberina-05-2005} Guberina, B., Horvat, R., \& Stefancic, H., Phys. Rev. D \textbf{72}, 025011 (2005).
\bibitem{guberina-05-2006} Guberina, B., Horvat, R., \& Nikolic, P., Phys. Lett. B \textbf{636}, 80 (2006).
\bibitem{gong-09-2004} Gong, Y. G., Phys. Rev. D \textbf{70}, 064029 (2004).
\bibitem{jamil-01-2010} Jamil, M., Saridakis, E. N., \& Setare, M. R., Phys. Lett. B \textbf{679}, 172 (2009).
\bibitem{J2011} Jamil, M., \& Sheykhi, A., Int. J. Theor. Phys. \textbf{50}, 625 (2011).
\bibitem{SJ2011} Sheykhi, A., \& Jamil, M., Gen. Rel. Grav. \textbf{43}, 2661 (2011).
\bibitem{she2011} Sheykhi, A., \& Jamil, M., Phys. Lett. B \textbf{694}, 284 (2011).
\bibitem{sheykhi-01-2011} Sheykhi, A., \& Jamil, M., Gen. Relativ. Gravit. \textbf{43}, 2661 (2011).
\bibitem{sheykhi-03-01-2010} Sheykhi, A., Class. Quantum Grav. \textbf{27}, 025007 (2010).
\bibitem{elizalde-05-2005} Elizalde, E., Nojiri, S., Odintsov, S. D., \& Wang, P., Phys. Rev. D \textbf{71}, 103504 (2005).
\bibitem{setare-02-2010} Setare, M. R., \& Jamil, M., JCAP \textbf{1002}, 010 (2010).
\bibitem{2011setare} Setare, M. R., \& Jamil, M., Europhys. Lett. \textbf{92}, 49003 (2010).
\bibitem{2010setarejamil} Setare, M. R., \& Jamil, M., Phys. Lett. B \textbf{690}, 1 (2010).
\bibitem{2011karami} Karami, K., et al., Gen. Relativ. Gravit. \textbf{43}, 27 (2011).
\bibitem{2010farooq} Farooq, M. U., Jamil, M., \& Rashid, M. A., Int. J. Theor. Phys. \textbf{49}, 2278 (2010).
\bibitem{SKJ2010} Sheykhi, A., Karami, K., \& Jamil, M., arXiv:1012.4345.
\bibitem{2010IJSF} Jamil, M., \& Farooq, M. U., Int. J. Theor. Phys. \textbf{49}, 42 (2010).
\bibitem{2009JSS} Jamil, M., et al., J. Korean Phys. Soc. \textbf{57}, 1137 (2010).
\bibitem{karami-03-2010} Karami, K., \& Sorouri, A., Phys. Scripta \textbf{82}, 025901 (2010).
\bibitem{sheykhi-11-2009} Sheykhi, A., Phys. Lett. B \textbf{681}, 205 (2009).
\bibitem{setare-09-2006} Setare, M. R., Phys. Lett. B \textbf{642}, 421 (2006).
\bibitem{setare-08-2008} Setare, M. R., \& Saridakis, E. N., Phys. Lett. B \textbf{668}, 177 (2008).
\bibitem{zhangX-08-2005} Zhang, X., \& Wu, F. Q., Phys. Rev. D \textbf{72}, 043524 (2005).
\bibitem{zhangX-11-2006} Zhang, X., \& Wu, F. Q., Phys. Rev. D \textbf{76}, 023502 (2007).
\bibitem{zhangX-07-2007} Zhang, X., Phys. Lett. B \textbf{648}, 1 (2007).
\bibitem{enqvist-02-2005} Enqvist, K., Hannestad, S., \& Sloth, M. S., JCAP \textbf{0502}, 004 (2005).
\bibitem{shen} Shen, J., Wang, B., Abdalla, E., \& Su, R. K., Phys. Lett. B \textbf{609}, 200 (2005).
\bibitem{jamilfarooq2009} Jamil, M., Farooq, M. U., \& Asif, M., Eur. Phys. J. C \textbf{61}, 471 (2009).
\bibitem{wangY} Wang, Y., \& Xu, L., Phys. Rev. D \textbf{81}, 083523 (2010).
\bibitem{micheletti} Micheletti, S. M. R., JCAP \textbf{1005}, 009 (2010).
\bibitem{zhangX-05-2009} Zhang, X., Phys. Rev. D \textbf{79}, 103409 (2009).
\bibitem{feng-02-2005} Feng, C. J., Phys. Lett. B \textbf{633}, 367 (2006).
\bibitem{kao} Kao, H. C., \& Lin, W. S., JCAP \textbf{0901}, 022 (2009).
\bibitem{huang-03-2005} Huang, Q. G., \& Li, M., JCAP \textbf{0503}, 001 (2005).
\bibitem{li-12-2004} Li, M., Phys. Lett. B \textbf{603}, 1 (2004).
\bibitem{pl1} Das, S., Shankaranarayanan, S., \& Sur, S., Phys. Rev. D \textbf{77}, 064013 (2008).
\bibitem{pl2} Radicella, N., \& Pavon, D., Phys. Lett. B \textbf{691}, 121 (2010).
\bibitem{Yassin:2020mjf} Yassin, Ahmed, et al., Int. J. Geom. Meth. Mod. Phys. \textbf{17}, 2050106 (2020).
\bibitem{Tawfik:2015fda} Tawfik, A. N., \& Diab, A. M., Int. J. Mod. Phys. D \textbf{24}, 1550085 (2015).
\bibitem{pl3} Sau, S., Shankaranarayanan, S., \& Das, S., Int. J. Mod. Phys. D \textbf{17}, 541 (2008).
\bibitem{banerjee-04-2008} Banerjee, R., \& Majhi, B. R., Phys. Lett. B \textbf{662}, 62 (2008).
\bibitem{banerjee-06-2008} Banerjee, R., \& Majhi, B. R., JHEP \textbf{0806}, 095 (2008).
\bibitem{banerjee-05-2009} Banerjee, R., \& Modak, S. K., JHEP \textbf{0905}, 063 (2009).
\bibitem{Majhi} Majhi, B. R., Phys. Rev. D \textbf{79}, 044005 (2009).
\bibitem{wei-02-2010} Wei, H., Commun. Theor. Phys. \textbf{52}, 743 (2009).
\bibitem{easson} Easson, D. A., Frampton, P. H., \& Smoot, G. F., Phys. Lett. B \textbf{696}, 273 (2011).
\bibitem{sad2010} Sadjadi, H. M., \& Jamil, M., Gen. Relativ. Gravit. \textbf{43}, 1759 (2011).
\bibitem{jam2011} Jamil, M., \& Sheykhi, A., Int. J. Theor. Phys. \textbf{50}, 625 (2011).
\bibitem{zhu} Zhu, T., \& Ji, J. R., JCAP \textbf{0911}, 007 (2009).
\bibitem{cai-08-2009} Cai, R. G., Cao, L. M., \& Hu, Y. P., Class. Quantum Grav. \textbf{26}, 155018 (2009).
\bibitem{nojiri-2001} Nojiri, S., \& Odintsov, S. D., Int. J. Mod. Phys. A \textbf{16}, 3273 (2001).
\bibitem{wei-10-2009} Wei, H., Commun. Theor. Phys. \textbf{52}, 743 (2009).
\bibitem{spergel-06-2007} Spergel, D. N., et al., Astrophys. J. Suppl. \textbf{170}, 377 (2007).
\bibitem{bertolami-10-2007} Bertolami, O., Gil Pedro, F., \& Le Delliou, M., Phys. Lett. B \textbf{654}, 165 (2007).
\bibitem{jamil-11-2008} Jamil, M., \& Rashid, M. A., Eur. Phys. J. C \textbf{58}, 111 (2008).
\bibitem{ichiki} Ichiki, K., \& Keum, Y. Y., JCAP \textbf{0806}, 005 (2008).
\bibitem{amendola-04-2007} Amendola, L., Campos, G. C., \& Rosenfeld, R., Phys. Rev. D \textbf{75}, 083506 (2007).

\bibitem{mioS1} Pasqua, A., Khodam-Mohammadi, A., Jamil, M., \& Myrzakulov, R., Astrophys. Space Sci. \textbf{340}, 199-208 (2012).
\bibitem{mioS2} Pasqua, A., Jamil, M., Myrzakulov, R., \& Majeed, B., Phys. Scr. \textbf{86}, 045004 (2012).
\bibitem{mioS3} Pasqua, A., Astrophys. Space Sci. \textbf{346}, 531–543 (2013).
\bibitem{mioS4} Pasqua, A., Chattopadhyay, S., Radinschi, I., Alshehri, A. A., \& Tawfik, A. N., Annals of Physics \textbf{465}, 169685 (2024).
\bibitem{mioS5} Pasqua, A. https://doi.org/10.48550/arXiv.2509.08029
\bibitem{mioS6} Pasqua, A.  https://doi.org/10.48550/arXiv.2509.19386
\bibitem{mioS7} Pasqua, A.  https://doi.org/10.48550/arXiv.2510.17488



\bibitem{bagla} Bagla, J. S., Jassal, H. K., \& Padmanabhan, T., Phys. Rev. D \textbf{67}, 063504 (2003).
\bibitem{shao} Shao, Y. S., et al., Mod. Phys. Lett. A \textbf{22}, 1175 (2007).
\bibitem{calcagni} Calcagni, G., \& Liddle, A. R., Phys. Rev. D \textbf{74}, 043528 (2006).
\bibitem{copeland-02-2005} Copeland, E. J., et al., Phys. Rev. D \textbf{71}, 043513 (2005).
\bibitem{sen-07-2002} Sen, A., JHEP \textbf{0207}, 065 (2002).
\bibitem{sen-10-1999} Sen, A., JHEP \textbf{9910}, 008 (1999).
\bibitem{bergshoeff} Bergshoeff, E., et al., JHEP \textbf{0005}, 009 (2000).
\bibitem{klusovn} Kluson, J., Phys. Rev. D \textbf{62}, 126003 (2000).
\bibitem{kutasov} Kutasov, D., Marinescu, M., \& Rey, S. J., Phys. Rev. D \textbf{55}, 6222 (1997).


\bibitem{sen-2002} Sen, A., Modern Physics Letters A \textbf{17}, 1797 (2002).
\bibitem{lambert} Lambert, N.~D., Sachs, I., Phys. Rev. D \textbf{67}, 025005 (2003).
\bibitem{gcg1} Kamenshchik, A., Moschella, U., Pasquier, V., Phys. Lett. B \textbf{511}, 265 (2001).
\bibitem{gcg6} Li, M., Li, X., Zhang, X., Science in China G: Physics and Astronomy \textbf{53}, 1631 (2010).
\bibitem{gcg9} Bento, M.~C., Bertolami, O., Sen, A.~A., Phys. Rev. D \textbf{70}, 083519 (2004).
\bibitem{gcg16} Gorini, V., Kamenshchik, A., Moschella, U., Phys. Rev. D \textbf{67}, 063509 (2003).
\bibitem{gcg17} Alam, U., Sahni, V., Saini, T.~D., Starobinsky, A.~A., Mon. Not. R. Astron. Soc. \textbf{344}, 1057 (2003).
\bibitem{bento} Bento, M.~C., Bertolami, O., Sen, A.~A., Phys. Rev. D \textbf{66}, 043507 (2002).
\bibitem{gcg4} Bilic, N., Tupper, G.~B., Viollier, R.~D., Phys. Lett. B \textbf{535}, 17 (2002).
\bibitem{gcg5} Fabris, J.~C., Goncalves, S.~B.~V., de Souza, P.~E., Gen. Rel. Grav. \textbf{34}, 53 (2002).
\bibitem{gcg15} Gorini, V., Kamenshchik, A.~Y., Moschella, U., Piattella, O.~F., Starobinsky, A.~A., J. Cosmol. Astropart. Phys. \textbf{02}, 016 (2008).
\bibitem{gcg7} Zhang, X., Zhang, J., Cui, J., Zhang, L., Mod. Phys. Lett. A \textbf{24}, 1763 (2009).
\bibitem{ym1} Zhang, Y., Phys. Lett. B \textbf{340}, 18 (1994).
\bibitem{ym9-3} Tong, M., Zhang, Y., \& Xia, T., Int. J. Mod. Phys. D \textbf{18}, 797 (2009).
\bibitem{ym9-5} Zhao, W., Astron. Astrophys. \textbf{9}, 874 (2009).
\bibitem{ym9-7} Zhang, Y., Xia, T. Y., \& Zhao, W., Class. Quantum Grav. \textbf{24}, 3309 (2007).
\bibitem{ym18-1} Gross, D. J., \& Wilzcek, F., Phys. Rev. Lett. \textbf{30}, 1343 (1973).
\bibitem{ym18} Pagels, H., \& Tomboulis, E., Nucl. Phys. B \textbf{143}, 485 (1978).
\bibitem{ym21} Adler, S. L., Phys. Rev. D \textbf{23}, 2905 (1981).
\bibitem{ym21-1} Adler, S. L., Nucl. Phys. B \textbf{217}, 388 (1983).
\bibitem{ym19} Coleman, S., \& Weinberg, E., Phys. Rev. D \textbf{7}, 1888 (1973). 
\bibitem{ym20} Parker, L., \& Raval, A., Phys. Rev. D \textbf{60}, 063512 (1999).
\bibitem{ym10-1} Barreiro, T., Copeland, E. J., \& Nunes, N. J., Phys. Rev. D \textbf{61}, 127301 (2002).
\bibitem{ele1} Breton, N., \& Garcia-Salcedo, R., arXiv:hep-th/0702008 (2007).
\bibitem{ele2} De Lorenci, V. A., Klippert, R., Novello, M., \& Salim, J. M., Phys. Rev. D \textbf{65}, 063501 (2002).


\end{thebibliography}
\end{document}